\documentclass[11pt]{article}
\usepackage{jcappub}
\usepackage{array, natbib, amsfonts}

\newcommand{\hMpc}{{$h^{-1}$Mpc}}
\newcommand{\hMpcinv}{{$h\,{\rm Mpc}^{-1}$}}
\newcommand{\grms}{{$\gamma_{\rm rms}$}}
\newcommand{\mnras}{{\rm MNRAS}}
\newcommand{\apj}{{\rm ApJ}}
\newcommand{\apjs}{{\rm ApJS}}
\newcommand{\araa}{{\rm Annu.Rev.Astron.Astrophys.}}

\newcommand{\aap}{{\rm Astron.Astrophys.}}
\newcommand{\nat}{{\rm Nature}}
\newcommand{\prd}{{\rm Phys.Rev.D}}
\newcommand{\apjl}{{\rm ApJL}}
\newcommand{\aj}{{\rm AJ}}

\title{Testing the tidal alignment model of galaxy intrinsic alignment}
\author[a]{Jonathan Blazek,}
\author[b]{Matthew McQuinn,}
\author[a,b,c,d]{and Uro\v{s} Seljak}

\affiliation[a]{Department of Physics and Lawrence Berkeley National Laboratory,\newline University of California, Berkeley, CA 94720, U.S.A.}
\affiliation[b]{Department of Astronomy, University of California, Berkeley, CA 94720, U.S.A.}
\affiliation[c]{Institute of Theoretical Physics, University of Zurich, Winterthurerstrasse 190,\newline CH-8057, Zurich, Switzerland}
\affiliation[d]{Institute for the Early Universe, Ewha University, Seoul 120-750, South Korea}

\emailAdd{blazek@berkeley.edu}

\abstract{Weak gravitational lensing has become a powerful probe of large-scale structure and cosmological parameters.  Precision weak lensing measurements require an understanding of the intrinsic alignment of galaxy ellipticities, which can in turn inform models of galaxy formation.  It is hypothesized that elliptical galaxies align with the background tidal field and that this alignment mechanism dominates the correlation between ellipticities on cosmological scales (in the absence of lensing).  We use recent large-scale structure measurements from the Sloan Digital Sky Survey to test this picture with several statistics: (1) the correlation between ellipticity and galaxy overdensity, $w_{g+}$; (2) the intrinsic alignment auto-correlation functions; (3) the correlation functions of curl-free, $E$, and divergence-free, $B$, modes, the latter of which is zero in the linear tidal alignment theory; (4) the alignment correlation function, $w_g(r_p,\theta)$, a recently developed statistic that generalizes the galaxy correlation function to account for the angle between the galaxy separation vector and the principle axis of ellipticity.  We show that recent measurements are largely consistent with the tidal alignment model and discuss dependence on galaxy luminosity.  In addition, we show that at linear order the tidal alignment model predicts that the angular dependence of $w_g(r_p,\theta)$ is simply $w_{g+}(r_p) \cos(2 \theta)$ and that this dependence is consistent with recent measurements.  We also study how stochastic nonlinear contributions to galaxy ellipticity impact these statistics.  We find that a significant fraction of the observed LRG ellipticity can be explained by alignment with the tidal field on scales $\gtrsim$ 10 \hMpc.  These considerations are relevant to galaxy formation and evolution.}

\arxivnumber{}

\begin{document}
\maketitle
\flushbottom

\section{Introduction}

Light travels through the web of cosmic structure on geodesic paths determined by the presence of matter.  The gravitational deflections due to matter inhomogeneities result in distorted observations of distant objects.  In special cases, the distortions can be profound, producing multiple images and even Einstein rings.  More commonly, the distortions are subtle and yield only small deviations from the intrinsic shape of the source, a process termed ``weak lensing."  The statistical analysis of these small distortions is becoming an important tool in the study of large-scale structure and cosmological parameters \cite{blandford91,miralda91,kaiser92,refregier03}.  Since lensing arises from purely gravitational physics, it directly probes the underlying matter rather than an observable that only correlates with some fraction of the matter.  After early successes in detecting the weak lensing signal \cite{bacon00,kaiser00,vanWaerbeke00,wittman00}, surveys such as COSMOS\footnote{http://cosmos.astro.caltech.edu/; \cite{scoville07}} and CFHTLS\footnote{Canada France Hawaii Telescope Legacy Survey, http://www.cfht.hawaii.edu/Science/CFHLS} have allowed more precise measurements on larger scales.  Among other things, weak lensing surveys can test General Relativity \cite{reyes10,daniel10,thomas09,lombriser10}, constrain the properties of dark energy \cite{albrecht06}, and measure galaxy bias \cite{simon07}.  Planned surveys such as DES\footnote{Dark Energy Survey, https://www.darkenergysurvey.org}, LSST\footnote{Large Synoptic Survey Telescope, http://www.lsst.org}, WFIRST\footnote{Wide-Field Infrared Survey Telescope, http://wfirst.gsfc.nasa.gov/}, Euclid\footnote{http://sci.esa.int/science-e/www/area/index.cfm?fareaid=102}, and Pan-STARRS\footnote{Panoramic Survey Telescope \& Rapid Response System, http://pan-starrs.ifa.hawaii.edu} have the potential to further improve the precision of weak lensing measurements.

However, these measurements are far from straightforward, requiring an understanding of a wide variety of systematic errors (see, e.g., \cite{mandelbaum06a} and references therein).  The past several years have seen a large effort to better understand and correct for instrumental systematics. However, even for a perfectly understood instrument, astrophysical uncertainties can contribute substantial errors when inferring the matter distribution from weak lensing measurements.  For instance, the intrinsic orientations and shapes of observed galaxies are a source of systematic error.  This ``intrinsic alignment'' (IA) of galaxies is particularly problematic since it can be significant and would bias even an ideal measurement.  Weak lensing statistics involve averaging across pairs of observed galaxy ellipticities $\gamma^{\rm obs}$, which consist of both the intrinsic ellipticity (I) of the galaxy and the gravitational lensing shear distortion (G): $\gamma^{\rm obs} = \gamma^{\rm I}+\gamma^{\rm G}$.  For the $i$-component of ellipticity, the observed ellipticity correlation function is then
\begin{align}
\label{a}
\langle \gamma_i^{\rm obs} \gamma_i^{\rm obs}\rangle = \langle \gamma_i^{\rm G} \gamma_i^{\rm G}\rangle + \langle \gamma_i^{\rm I} \gamma_i^{\rm G}\rangle + \langle \gamma_i^{\rm G} \gamma_i^{\rm I} \rangle + \langle \gamma_i^{\rm I} \gamma_i^{\rm I}\rangle.
\end{align}

The desired signal in weak lensing studies is the first term (GG).  If intrinsic alignments are random, the GI and II terms average to zero.  However, the coherent influence of large-scale structure on galaxy ellipticity contaminates the lensing signal because intrinsic shapes acquire a non-zero average correlation.  Since the weak lensing signal is small ($\gamma^{\rm G}$ is roughly 1\% of $\gamma^{\rm I}$ for a typical galaxy \cite{hirata04}), even small correlations can lead to appreciable intrinsic alignment contributions.  It has long been known that the II correlations from galaxies in close proximity (where intrinsic ellipticity correlations are strongest) could be a significant contaminant \cite{croft00,heavens00,catelan01,crittenden01,jing02}.  Fortunately, the II term can be easily reduced by either down-weighting or excluding nearby pairs \cite{king02,king03,heymans03,takada04}.  It was later realized by \cite{hirata04} that the GI term can also be a significant contaminant, introducing a correlation in the ellipticities of objects that are along the same line-of-sight but separated by a large spatial distance.  A foreground lensing potential affects the intrinsic ellipticities of nearby objects as well as the observed ellipticites of background objects via lensing.  Observations have confirmed the presence of both of these intrinsic effects \cite{mandelbaum06a,hirata07,faltenbacher09,okumura09a,okumura09b}.

It is critical to understand IA for high-precision weak lensing experiments.  In \cite{hirata07}, it was shown that GI contamination at the $\sim 10\%$ level is possible in a typical weak lensing survey.  The potential degradation of cosmological parameter measurements by IA contamination is significant.  For instance, IA can bias cosmic shear measurements of $\sigma_8$ at the current level of uncertainty ($\sigma_8\approx 0.8\pm0.07$) for a CFHLTS-like survey, which is larger by $\approx$ 3 than the best current constraints \cite{schneider10}.  Similarly, uncertainty in the amplitude of intrinsic alignments can impart a significant bias in cosmological parameter measurements, even when a particular model is assumed in order to subtract the alignment signal \cite{joachimi10}.  However, these alignment effects are not just a contaminant -- they also provide a probe of large-scale structure and galaxy formation.

Several models of IA with varying levels of complexity have been proposed (e.g.~\cite{catelan01,lee08a,schneider10,hui02}).  These models belong to two general classes: alignment of the galaxies with background tidal field or torquing of the galactic angular momentum vector by the tidal field.  Any analytic predictions of the orientation and ellipticity of a galaxy residing in a background tidal field relies on assumptions relating the orientation of dark matter halos with that of the resident galaxies, and nonlinear scales are particularly difficult to model analytically.  Intrinsic alignment of dark matter halos has also been studied using $N$-body simulations \cite{ciotti94,ciotti98,pereira08,pereira10,faltenbacher09}, which have shown that ellipticities can differ significantly in both shape and orientation when measured in the inner and outer regions of halos.  A better understanding of the relationship between halo and galaxy ellipticities may require simulations with baryon physics and an exceptional dynamic range (see, e.g., \cite{hahn10}).

This paper tests analytic models and different statistical measures.  In particular, we focus on the linear tidal alignment (LA) model \cite{catelan01,hirata04}, which posits that the intrinsic ellipticity of a galaxy is a linear function of the tidal field.  This model should dominate on large scales for elliptical galaxies.  Recent work by \cite{joachimi10} has shown that the model is consistent with measurements of GI correlation.  We expand this comparison and consider possible extensions to the model.

This paper is organized as follows.  Section \ref{sec:model} discusses intrinsic alignment models, especially the linear tidal alignment model.  In Section \ref{sec:statistics}, we summarize recent measurements of intrinsic alignment.  We then review several statistics relevant for GI and II correlations and calculate these statistics in the LA model. One of these statistics is the alignment correlation function, $w_g(r_p,\theta)$, recently proposed in \cite{faltenbacher09}.  By adding an angular dimension to the correlation function, this statistic can in principle contain additional information on the relationship between galaxy clustering and alignment, a prospect which we examine. We compare the model predictions to recent measurements and determine the consistency and strength of linear tidal alignment.  We also propose a potential signature of nonlinear alignment contributions and nonlinearities in the density field.  In Section \ref{sec:stoc}, we consider the effects of a stochastic contribution to galaxy ellipticity that does not correlate between galaxies.  We study two models for this stochastic component and discuss the impact on the measured alignment statistics.  Section \ref{sec:summ} summarizes our conclusions and provides discussion in the context of galaxy formation.  We also include an appendix with the details of some calculations referenced in the text.  Throughout this work, we assume a flat $\Lambda$CDM cosmology with $\Omega_m=0.25$, $\Omega_b=0.04$, $\sigma_8=0.8$, $n_s=1$, and $H_0=70~{\rm km~s^{-1}~Mpc^{-1}}$.

\section{The linear tidal alignment model}
\label{sec:model}
The collapse of overdense regions into dark matter halos and galaxies occurs preferentially along the stretching axis of a background tidal field, and galaxy intrinsic ellipticity should maintain some memory of this asymmetry at the time of formation \cite{catelan01}.  In particular, elliptical galaxies are supported primarily through velocity dispersion rather than rotation and are thus more likely to align with the surrounding halo and background tidal field.  It is physically reasonable for galaxy orientation to correlate with the principal axis of the gravitational tidal field \cite{hirata04}.  The LA model of \cite{catelan01} relates the intrinsic ellipticity\footnote{The quantity used here is actually the intrinsic shear, which differs from ellipticity by a factor $1/2R$, described after eq.~\ref{c}.  To avoid confusion with lensing shear, we refer only to intrinsic ellipticity.} of an elliptical galaxy to a linear function of the tidal field:
\begin{align} 
\label{b}
\gamma^I_{(+,\times)}=-\frac{C_1}{4\pi G}(\nabla_x^2-\nabla_y^2,2\nabla_x\nabla_y)S[\Psi_P],
\end{align}
where $C_1$ parameterizes the strength of the alignment, with sign convention such that positive $C_1$ corresponds to preferential galaxy alignment along the stretching axis of the tidal field.\footnote{We define $C_1$ to capture the full magnitude of the LA effect.  In several previous studies (e.g.~\cite{joachimi10}), $C_1$ was specified using a standard but somewhat arbitrary normalization calculated from ellipticity variance \cite{hirata04}, and an additional dimensionless constant parameterized the strength of LA with respect to this reference value.  As we discuss in section \ref{sec:stoc}, stochastic contributions can affect large-scale correlations differently than ellipticity variance.  Since the LA model is most applicable on large scales, we choose a convention for $C_1$ that relates it directly to the magnitude of large-scale correlations.}  Positive $C_1$ thus yields an anti-correlation between the intrinsic alignment of a foreground object and the gravitational shear of a background object.  $\Psi_P$ is the gravitational potential, and $S$ is a filter that smooths fluctuations on halo scales. In \cite{hirata04}, $S$ is chosen to be a top-hat in Fourier space with a maximum wavevector of 1 \hMpcinv. When compared with no smoothing, we find that this choice has a negligible effect on scales of interest ($\gtrsim 10$ \hMpc), and we thus effectively ignore $S$. Up to derivatives, eq. \eqref{b} is the unique function of $\Psi_P$ that is local, linear, and quadrupole symmetric.  Since higher-derivative terms should be negligible on large scales, the LA model is unique up to the normalization $C_1$.  The $x$- and $y$-axes in eq. \eqref{b} are on the plane of the sky, and ellipticity is decomposed with respect to this coordinate system:
\begin{align}
\label{c}
\left[\begin{array}{c}
\gamma_+ \\
\gamma_{\times} \end{array}
\right]
=
 \left(\frac{1}{2R}\right)\left(\frac{1-(b/a)^2}{1+(b/a)^2}\right)
\left[\begin{array}{c}
\cos(2\phi) \\
\sin(2\phi)\end{array}
\right]
\equiv
\gamma_0
\left[\begin{array}{c}
\cos(2\phi) \\
\sin(2\phi)\end{array}
\right]
\end{align}
where $\phi$ is the position angle measured from the $x$-axis, and $b/a$ is the axis ratio.  $R$ is the shear responsivity, which captures the average response of measured ellipticity to a small shear \cite{bernstein02}.

The density-weighted intrinsic ellipticity is defined as $\tilde{\gamma}^I \equiv (1+\delta_g)\gamma^I$, for galaxy overdensity $\delta_g = b_g \delta$, where $\delta$ is the matter overdensity, and $b_g$ is the galaxy linear bias factor.  In the linear regime, the gravitational potential at redshift $z_p$ is related to the density field via the Poisson equation (valid on sub-horizon scales, $k\gg cH_0$):
\begin{align}
\label{f}
\Psi_P(\boldsymbol{k}) = -4\pi G \rho_{m,0}(1+z_P) k^{-2}\delta(\boldsymbol{k},z_P),
\end{align}
where $\rho_{{\rm m},0}$ is the present mean matter density.  There is some ambiguity to the appropriate redshift, $z_P$, to evaluate the potential.  In previous work \cite{hirata04,joachimi10}, $z_P$ was chosen to be during matter domination, when most of the stars in elliptical galaxies formed.  However, it is also plausible that recent accretion significantly impacts the alignment, in which case the gravitational potential should be evaluated at roughly the observed redshift.  In linear theory, the only difference between the potential at different times is the overall amplitude, which can be absorbed into $C_1$.  Thus, the inferred strength of IA will depend on $z_P$.  For consistency with previous work, we evaluate $\Psi_P$ during matter domination.

A product of density fields in configuration space becomes a convolution in Fourier space, and thus the linear model predicts:
\begin{align}
\label{g}
\tilde{\gamma}^I_{(+,\times)}(\boldsymbol{k},z) = \frac{-C_1\rho_{m,0}}{D(z)}\int d^3\boldsymbol{k_1}\frac{(k^2_{2x}-k^2_{2y},2k_{2x}k_{2y})}{k^2_2}\delta(\boldsymbol{k_2},z)\left[\delta^{(3)}(\boldsymbol{k_1})+\frac{b_g}{(2\pi)^3}\delta(\boldsymbol{k_1},z)\right],
\end{align}
where $\boldsymbol{k_2} \equiv \boldsymbol{k} - \boldsymbol{k_1}$, $\delta^{(3)}(\boldsymbol{k})$ denotes the 3-dimensional Dirac delta function, and $D(z)$ is the growth factor, normalized so that $(1+z)D(z)=1$ during matter domination.

The LA model breaks down on nonlinear scales.  As in recent work \cite{joachimi10,schneider10}, we estimate these effects by using the Halofit nonlinear density power spectrum \cite{smith2003}.  We note, however, that simply applying a nonlinear power spectrum to the LA model is not a full nonlinear theory.  Moreover, the resulting nonlinear corrections depend on the choice of $z_P$, since the growth function does not fully capture the evolution of the nonlinear density field.  For instance, choosing $z_P$ at the observed redshift (which provides the maximum nonlinear correction) rather than during matter domination can affect the predicted alignment amplitude by $\sim 20 \%$ for GI correlations at scales of $5$\hMpc.

On large scales, the LA mechanism should dominate ellipticity correlations, which will scale linearly with the matter power spectrum, $P_{\delta}(k)$.  Spiral galaxies are supported by angular momentum, and thus a distinct alignment mechanism, based on the tidal torquing theory of protogalaxies, may be relevant.  Models based on tidal torquing can be categorized as ``quadratic alignment models," since the tidal field enters quadratically at lowest order rather than linearly \cite{pen00,catelan01,hui02}, suppressing large-scale correlations because $\delta \ll 1$.  Predictions of intrinsic alignment effects from quadratic models are qualitatively different from the linear model.  For example, quadratic models predict a divergence-free ($B$-mode) component to the ellipticity at leading order \cite{mackey02} but a vanishing lowest-order correlation between matter density and ellipticity.  Nonlinearities in the density field could potentially allow quadratic alignment effects to contribute at linear order in $P_{\delta}(k)$ \cite{hui02}.  Recent observations \cite{faltenbacher09,hirata07,mandelbaum10} have split galaxies by color into ``red" and ``blue" sub-samples, finding qualitative differences in intrinsic alignment, suggesting the possibility of different alignment mechanisms.  Blue samples exhibit weaker intrinsic alignment on large scales, supporting the theory that LA effects are less prominent in spirals.

\section{Measuring intrinsic alignment}
\label{sec:statistics}
There are numerous probes of galaxy intrinsic alignment.  We consider several alignment statistics in real space (for both GI and II correlations) and compare measurements with LA model predictions.  Table \ref{tab:summary} provides a summary of these statistics, which are described in the following subsections.

 \begin{table*}[t!]
\begin{center}
\begin{tabular}{|c|c|c|c|c|}
\hline
Test&$C_1 \rho_{\rm crit}$&$\chi^2_{\rm red.}$&$p(>\chi^2)$&Comments\\
\hline
$w_{g+}$&$0.125 \pm 0.007$&2.3&0.05&NL corrections improve fit below 10 \hMpc\\
\hline
$w_{++}$&$0.123 \pm 0.014$&0.43&0.79&NL corrections improve fit below 10 \hMpc\\
\hline
$w_{\times\times}$&Use $w_{++}$ fit&2.4&0.03&---\\
\hline
$w_{E}$&Use $w_{++}$ fit&2.8&0.02&---\\
\hline
$w_{B}$&---&0.68&0.64&LA prediction is $w_{B}=0$\\
\hline
&$\tilde{C}_1 \rho_{\rm crit}$&&&\\
\hline
$\tilde{w}_{g+}$&$0.71 \pm 0.02$&1.8&0.12&Calculated without weighting by $\gamma_0$\\
\hline
$\tilde{w}_{++}$&$0.74 \pm 0.07$&0.24&0.91&Calculated without weighting by $\gamma_0$\\
\hline
$w_g(r_p,\theta)$&$0.16-1.55$&---&---&Luminosity dependent - see section \ref{sec:align}\\
\hline
\end{tabular}
\end{center}
\caption{Summary of tests of LA model, including measured model parameter and reduced $\chi^2$ for the fit.  All measurements use the SDSS LRG catalog except for $w_g(r_p,\theta)$, which uses the SDSS main sample.
The upper section contains statistics that are weighted by the ellipticity magnitude while the lower section contains unweighted statistics.}
\label{tab:summary}
\end{table*}

\subsection{Galaxy samples}
To test the predictions of the linear alignment model, we compare with existing measurements of intrinsic alignment statistics.  Catalogs from large, deep surveys have allowed recent measurements of these correlations with better precision and at larger separations than was previously possible.  In \cite{okumura09a,okumura09b}, the authors used the catalog of Luminous Red Galaxies (LRGs) from DR6 of the Sloan Digital Sky Survey (SDSS; \cite{york00}) to measure both II and GI correlations.  LRGs, which are among the most luminous elliptical galaxies, are expected to be particularly well described by the LA model.  The LRG sample used in their analysis contains 83,773 objects with spectroscopic redshifts in the range $0.16<z<0.47$ and mean redshift of $\bar{z}=0.32$.   Note that the LRG shapes used in \cite{okumura09a,okumura09b} are measured without correcting for the point-spread function (PSF).
We find that $\gamma_0$ is larger by $\approx$\;10\,-\,30\% when measured without PSF-correction (R. Mandelbaum, private communication).  This offset would affect the amplitude of any IA statistic that weights by ellipticity magnitude and thus the inferred value of $C_1$.

We also consider the measurements made in \cite{faltenbacher09} using SDSS DR6 main sample galaxies from the New York University Value Added Galaxy Catalog \cite{blanton05}.  This sample includes 430,164 galaxies with spectroscopic redshifts $0.01<z<0.40$. For more information on the redshift distribution of this sample, divided by luminosity, see \cite{mandelbaum05}.

Since we wish to test the LA model through comparison with observations of the ellipticity statistics, we face the opposite problem as cosmic shear experiments: weak lensing introduces a contaminating signal into the observations.  However, the lensing signal is negligible for the measurements employed here, which have low mean redshifts and where only correlations between galaxies within a projection volume of $\sim 100~$Mpc are included.

\subsection{Intrinsic alignment statistics}
Intrinsic alignment contributes both a cross-correlation (GI) term and an auto-correlation (II) term (see \cite{hirata04,mandelbaum06a,hirata07}).  We write the intrinsic ellipticity cross- and auto-correlation functions as:
\begin{align}
\label{d}
\xi_{gi} (\boldsymbol{r}) = \langle \delta_g(\boldsymbol{x}) \tilde{\gamma}^I_{i}(\boldsymbol{x+r})\rangle, \\
\xi_{ii} (\boldsymbol{r}) = \langle \tilde{\gamma}^I_{i}(\boldsymbol{x}) \tilde{\gamma}^I_{i}(\boldsymbol{x+r}) \rangle, \notag
\end{align}
where $i = \{+, \times \}$, and these two components are measured with respect to the separation vector on the sky.  Note that by symmetry, $\xi_{g\times} (\boldsymbol{r}) = 0$.  Since it is the background matter field that is responsible for gravitational lensing, $\xi_{g+}$ provides a measure of the GI correlation, while $\xi_{ii}$ probes the II correlations.

The separation vector $\boldsymbol{r}$ can be separated into a component on the sky ($r_p$) and along the line-of-sight ($\Pi$).  The projected correlation function of quantity $X$ can be written in terms of the 3D correlation function as
\begin{align}
\label{e}
w_{X} (r_p)= \int_{-\Pi_{{\rm max}}}^{\Pi_{{\rm max}}} \xi_{X}(r_p,\Pi) \,d\Pi,
\end{align}
where $2\,\Pi_{{\rm max}}$ is the depth of the projected volume.

It is straightforward to calculate these correlation functions in the LA model.  Unless otherwise specified, we include terms to quadratic order in $\delta$ in the following calculations.  In the LA model,
\begin{align}
\label{i}
\xi^{LA}_{g+} (r_p,\Pi) &= \langle \delta_g(\boldsymbol{0}) \tilde{\gamma}^I_+(r_p,\Pi)\rangle, \notag\\
\notag
&= \frac{-C_1 b_g \rho_{m,0}}{D(z)}\int \frac{d^3\boldsymbol{k}}{(2\pi)^3}\frac{d^3\boldsymbol{k'}}{(2\pi)^3}\langle\delta(\boldsymbol{k})\delta(\boldsymbol{k'})\rangle \left(\frac{k_x^2-k_y^2}{k^2}\right)e^{i(k_x r_p+k_z \Pi)}, \notag\\
&= \frac{C_1 b_g \rho_{m,0}}{2\pi^2 D(z)}\int d\kappa dk_z \frac{\kappa^3}{k^2}P_{\delta}(k,z)\cos(k_z \Pi)J_2(\kappa r_p).
\end{align}
We have defined the $x$-axis to be the separation axis on the sky and the $z$-axis to be along the line-of-sight.  $\kappa$ is the magnitude of the wavevector on the sky, while $k_z$ is the magnitude along the line-of-sight ($k^2 \equiv \kappa^2 + k_z^2$).  The projection of $\xi_{g+}^{LA}$ along the line-of-sight is
\begin{align}
\label{j}
w^{LA}_{g+}(r_p) = \frac{b_g}{\pi^2}\frac{C_1 \rho_{m,0}}{D(z)}\int d\kappa dk_z \frac{\kappa^3}{k^2 k_z}P_{\delta}(k,z)\sin(k_z \Pi_{\rm max})J_2(\kappa r_p).
\end{align}
Similarly, the projected auto-correlation statistics are given by
\begin{align}
\label{k}
w^{LA}_{(++,\times \times)}(r_p) = \frac{1}{{2\pi^2}} \left(\frac{C_1 \rho_{m,0}}{D(z)}\right)^2 \int d\kappa dk_z \frac{\kappa^5}{k^4 k_z}P_{\delta}(k,z) \sin(k_z \Pi_{\rm max})\left[J_0(\kappa r_p) \pm J_4(\kappa r_p)\right].
\end{align}

\begin{figure}[t!]
\begin{center}
\includegraphics[width=8.8cm]{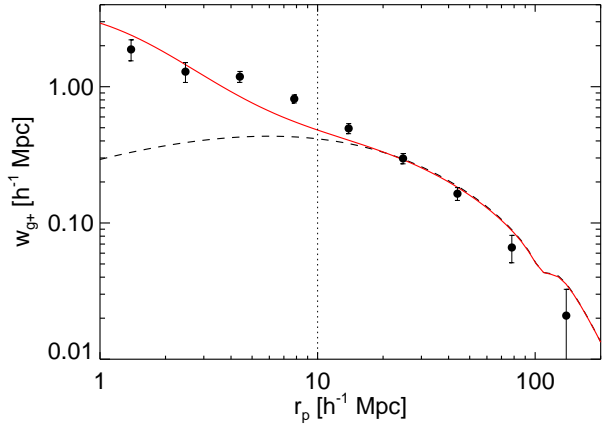}
\end{center}
\caption{Measurements of \protect\cite{okumura09b} and LA model prediction for $w_{g+}$.  The black dashed line is calculated using the linear theory $P_{\delta}(k)$, and the red solid line uses the Halofit model.}
\label{fig:wg+}
\end{figure}

\begin{figure}[t!]
\begin{center}
\resizebox{\hsize}{!}{
\includegraphics{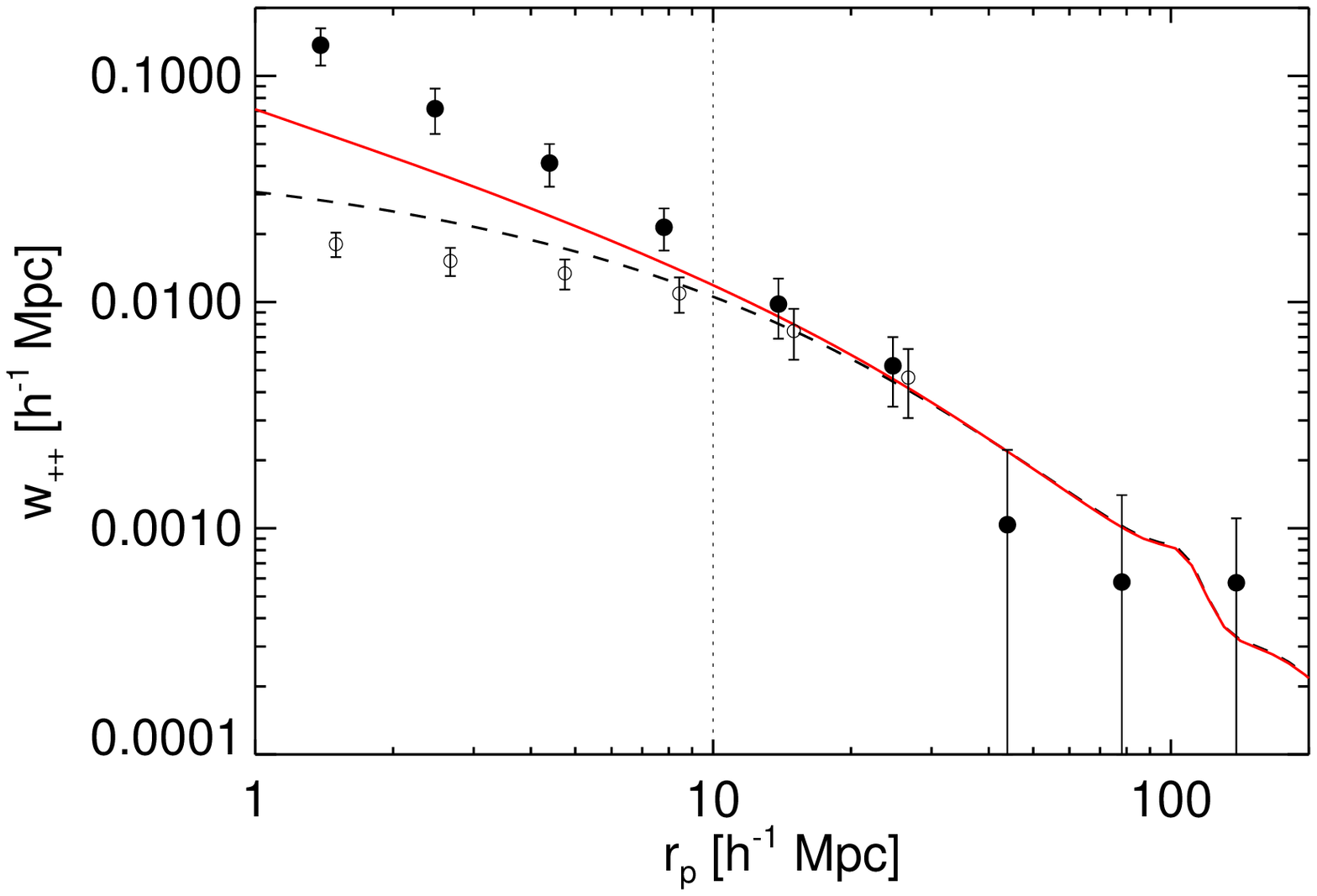}
\includegraphics{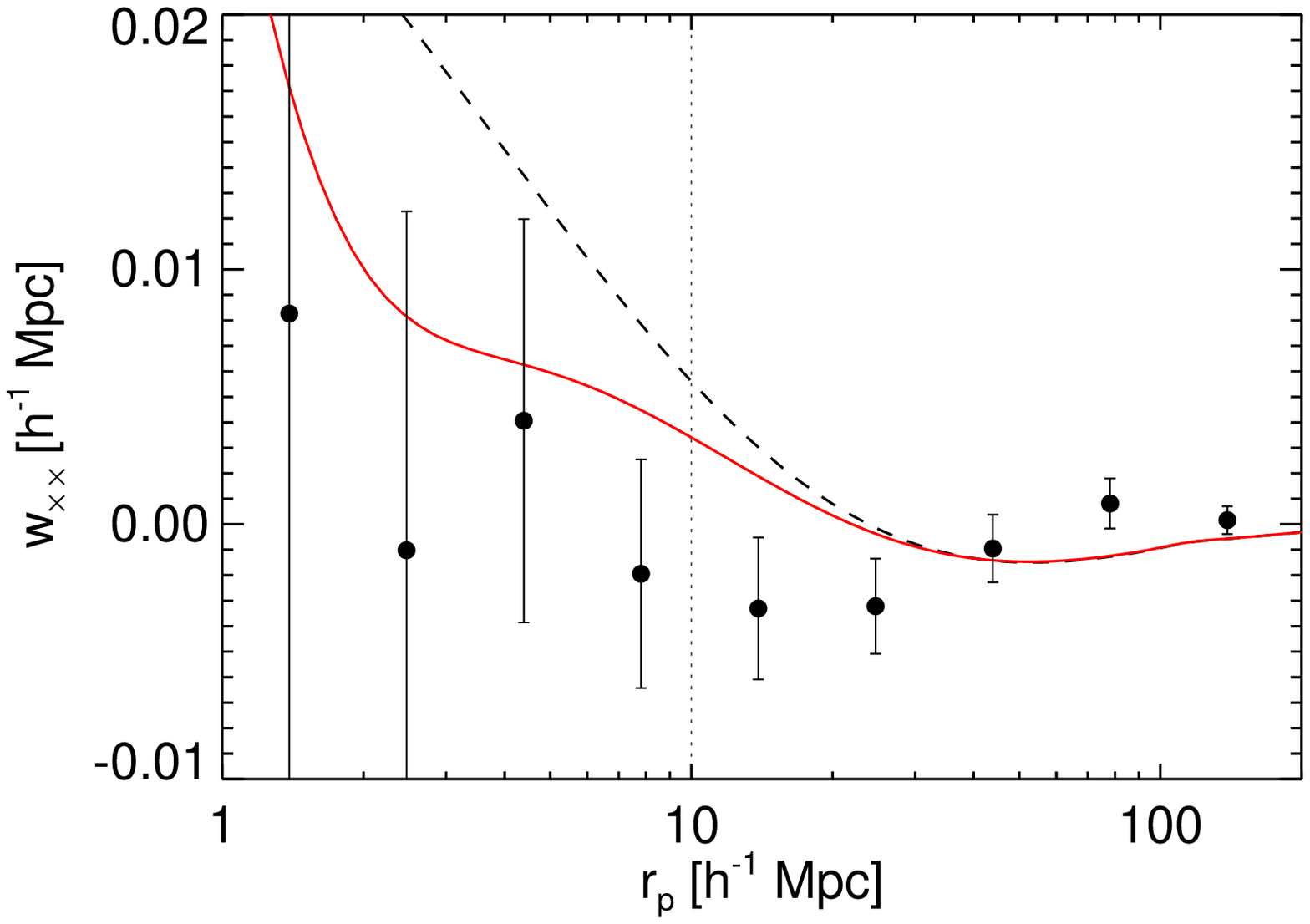}
}
\end{center}
\caption{Measurements of \protect\cite{okumura09a} and model predictions for $w_{++}$ ({\it left panel}) and $w_{\times \times}$ ({\it right panel}).  The measurements have been projected along the line-of-sight.  Open circles, indicating the original measurements without the $\left( 1+\xi_g(r)\right)$ correction, are only shown for $w_{++}$ and on small scales where there is an appreciable difference.  For clarity, these points have a small horizontal offset.  Line convention is the same as in figure~\ref{fig:wg+}.  A linear $y$-axis is used for $w_{\times \times}$.  The normalization of the LA prediction for both statistics is set from the fit to $w_{++}$.}
\label{fig:auto}
\end{figure}

In figures~\ref{fig:wg+}-\ref{fig:auto}, we compare the linear tidal alignment model predictions with measurements from SDSS LRGs \cite{okumura09a,okumura09b}, which project these statistics over $\Pi_{\rm max}=80$~\hMpc. Errors were calculated through jackknife resampling. Redshift-space distortions are included in the model predictions, as described in Appendix~\ref{app:4}.  We measure the galaxy bias $b_g$ by fitting to the measured projected galaxy correlation function, $w_{g}(r_p)$, and find $b^{\rm LRG}_g=2.12 \pm 0.04$.  The II correlations presented in \cite{okumura09a} are functions of 3-dimensional separation rather than projected separation.  To compare with the LA prediction, we have assumed that the statistics are isotropic and performed a projection along the line-of-sight following eq.~\eqref{e}.  To calculate the errors for these projections, we use the mean and variance of the 3-dimensional measurements and perform the projection on 1000 random realizations. In addition, the estimator used to calculate the II statistics in \cite{okumura09a} results in an additional factor of $\left( 1+\xi_g(r)\right)^{-1}$, where $\xi_g(r)$ is the 3-dimensional galaxy correlation function.  We have removed this factor before performing the projection.  For reference, the open circles in the left panel of figure~\ref{fig:auto} show the $w_{++}$ measurements without this correction.

To calculate the magnitude of $C_1$, we fit to the data at large separations ($r_p\approx$ 10\,-\,140\,\hMpc), where we expect the linear model to hold with minimal contamination from other alignment effects.  Below this scale, nonlinear effects become appreciable, and the LA model is no longer valid.  The LA theory agrees quite well with measurements at separations above $\approx$ 20 \hMpc\ for $w_{g+}$ (figure~\ref{fig:wg+}) and above $\approx$ 10 \hMpc\ for $w_{++}$ (figure~\ref{fig:auto}).  Including nonlinear corrections with Halofit improves agreement on smaller scales.  Agreement is weaker but still reasonable for the $w_{\times\times}$ statistic.

Table \ref{tab:summary} shows the fit results as well as reduced $\chi^2$ to indicate the goodness of fit.  We quote values for the dimensionless quantity $C_1 \rho_{\rm crit}$, where $\rho_{m,0}\equiv\Omega_m\rho_{\rm crit}$.  We fit separately to the II and GI statistics, both with and without ellipticity weighting.  Calculating these statistics without weighting by ellipticity is equivalent to setting the ellipticity magnitude, $\gamma_0$, of each galaxy equal to 1.  We define $\tilde{C}_1$, the non-weighted analog of $C_1$.  The ratio of weighted to non-weighted statistics is $\approx 0.17$, although as discussed below, some care must be taken to distinguish between this value and the mean ellipticity of the LRG sample.

We find that the $w_{++}$ and $w_{g+}$ statistics are well-described on large scales by the LA model with consistent amplitudes. The value of $C_1$ measured here is consistent with that found in \cite{joachimi10}, who examined $w_{g+}$ for a variety of data sets and found $C_1 \rho_{\rm crit} \approx 0.13 \pm0.02$ for a similar SDSS LRG sample (medium luminosity bin).\footnote{Note that \cite{joachimi10} use PSF-corrected galaxy shapes.}  

\subsection{Autocorrelation $E$- and $B$-modes}
The $w_{\times \times}$ and $w_{++}$ statistics can be written in terms of curl-free ($E$) and divergence-free ($B$) modes.  Lensing by matter produces only $E$-modes, making such a decomposition a useful diagnostic in studying the effects of intrinsic alignment and other systematics \cite{kamionkowski98}.  As shown below, only $E$-modes are produced in the LA model, and thus $B$-modes could indicate the presence of separate alignment mechanisms \cite{mackey02}.

Following \cite{crittenden02}, we can express the $E$- and $B$-components of the auto-correlation functions in real space in terms of the linear combinations $w_{\pm}(r_p) \equiv w_{++}(r_p) \pm w_{\times \times}(r_p)$:
\begin{align}
\label{l}
w_{(E,B)}(r_p) = \frac{w_+(r_p) \pm w'(r_p)}{2},
\end{align}
where $w'(r_p)$ is a non-local function of $w_-(r_p)$:
\begin{align}
\label{m}
w'(r_p) \equiv  w_-(r_p) + 4 \int_{r_p}^{\infty}dr' \frac{w_-(r')}{r'} -12r_p^2 \int_{r_p}^{\infty}dr' \frac{w_-(r')}{r'^3}.
\end{align}
From the integral properties and recursion relations of Bessel functions (see Appendix \ref{app:3}), $w'(r_p) = w_+(r_p)$ in the LA model, which allows us to write the $E/B$ decomposition:
\begin{align}
\label{n}
w^{LA}_E(r_p) &= \frac{1}{{\pi^2}} \left(\frac{C_1 \rho_{m,0}}{D(z)}\right)^2 \int d\kappa dk_z \frac{\kappa^5}{k^4 k_z}P_{\delta}(k)
\sin(k_z \Pi_{max})J_0(\kappa r_p), \notag\\
w^{LA}_B(r_p) &= 0.
\end{align}

\begin{figure}
\begin{center}
\resizebox{\hsize}{!}{
\includegraphics{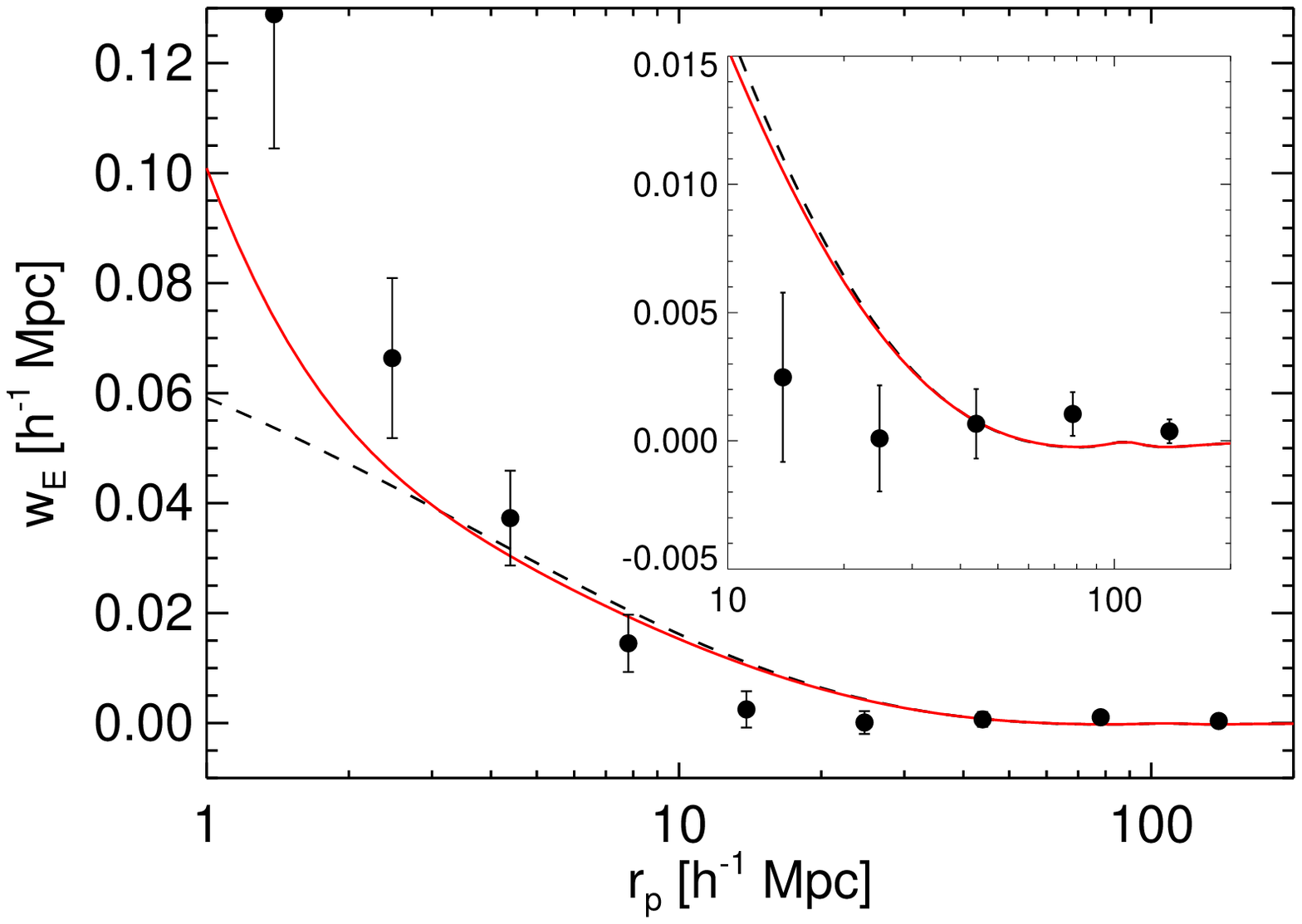}
\includegraphics{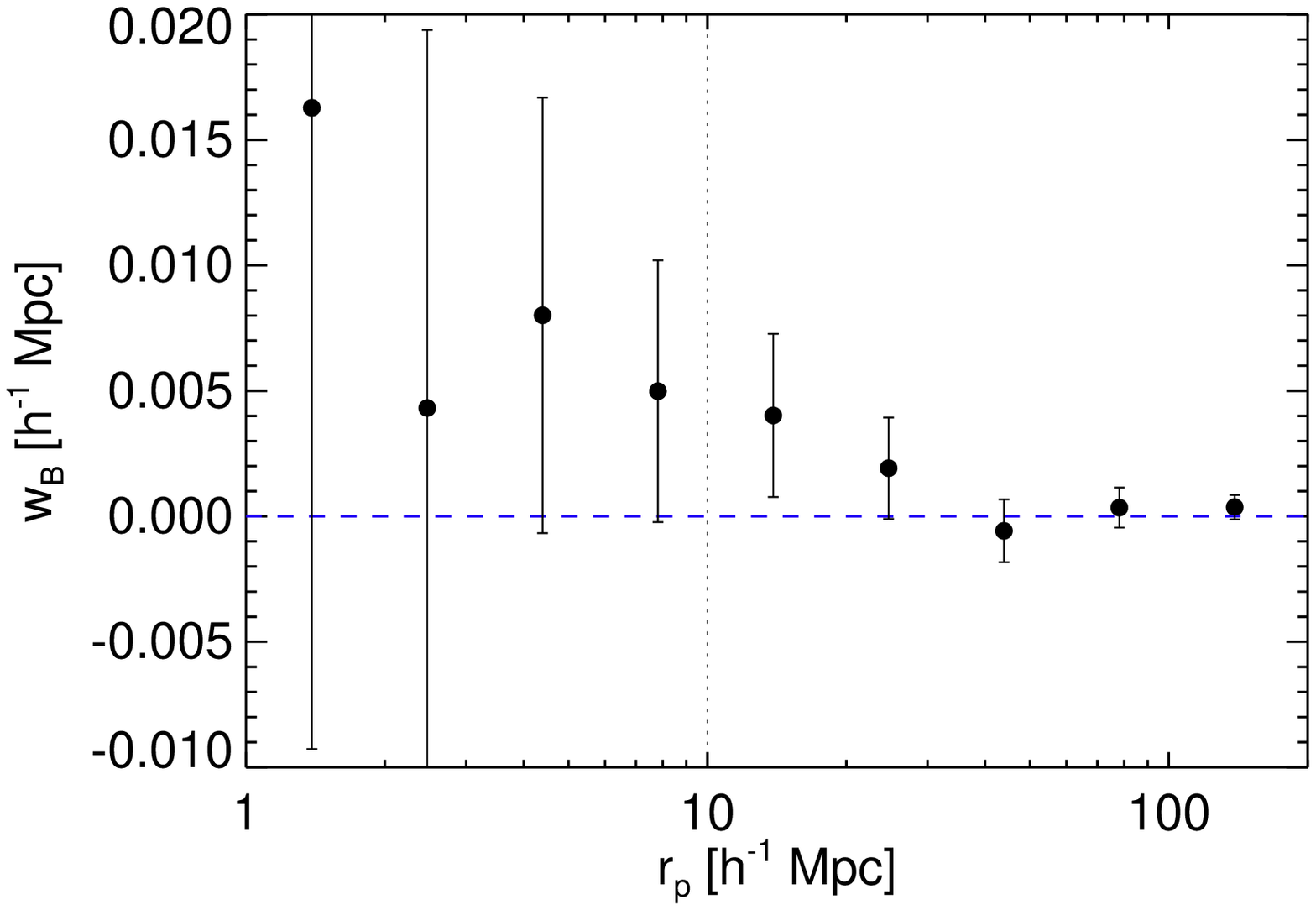}
}
\end{center}
\caption{{\it Left panel:} $E$-mode auto-correlation statistic.  Line convention and LA model normalization are the same as in figure~\ref{fig:auto}.  Inset shows more detail above $r_p=10$ \hMpc.  {\it Right panel:} $B$-mode auto-correlation statistic is compared with the LA prediction of zero.  The observations are consistent with the prediction above 10 \hMpc.}
\label{fig:autoEB}
\end{figure}

Projecting the auto-correlation measurements of SDSS LRGs and using eqs. \eqref{l} and \eqref{m}, we have calculated the observed $E$- and $B$-mode signals.  Errors are calculated as with $w_{++}$ and $w_{\times \times}$. Figure \ref{fig:autoEB} shows the LA predictions and measurements.  The observed $w_B(r_p)$ is consistent with the LA prediction of zero on scales above 10 \hMpc.  Below this scale, nonlinear terms become important, and it is not expected that $w_{B}$ would remain negligible.

\subsection{The alignment correlation function}
\label{sec:align}
Faltenbacher et al.~\cite{faltenbacher09} introduced the alignment correlation function, $w_{\rm g}(r_p,\theta)$, which describes the dependence of clustering on both projected separation, $r_p$, and the galaxy orientation angle, $\theta$, measured from the axis of separation.  It is related to the previously defined correlation statistics via the relations
\begin{align}
\label{v}
w_{\rm g}(r_p) = \frac{2}{\pi}\int_0^{\pi/2}d\theta \, w_{\rm g}(r_p,\theta); ~~~
\tilde{w}_{g+} = \frac{2}{\pi}\int_0^{\pi/2}{d\theta \,\cos(2\theta) \, w_{\rm g}(r_p,\theta)},
\end{align}
where $\tilde{w}_{g+} \equiv \langle \cos(2\phi) \delta_g\rangle$ is the $w_{g+}$ statistic when not weighted by ellipticity.  Because of its angular dependence, $w_{\rm g}(r_p,\theta)$ can in principle provide more information than $w_{g+}$.  Indeed, eq. \eqref{v} shows that $w_{g+}$ is the dipole moment of $w_{\rm g}(r_p,\theta)$ and thus lacks any information from higher angular moments.

We now derive $w_{\rm g}(r_p,\theta)$ in the LA model.  An arbitrary periodic function of $\theta$ can be written as a sum of $\cos(n\theta)$ and $\sin(n\theta)$ terms.  However, two symmetries exist for $w_g(r_p,\theta)$ which greatly restrict possible terms.  First, by parity, $w_g(r_p,\theta) = w_g(r_p,-\theta)$, ruling out all $\sin(n\theta)$ terms.  Second, since the ellipticities are invariant under rotations by $\pi$, $w_g(r_p,\theta) = w_g(r_p,\theta + \pi)$, which allows only $\cos(n\theta)$ terms with $n$ even.  In its most general form, therefore,
\begin{align}
\label{p}
w_{\rm g}(r_p,\theta) = w_{\rm g}(r_p) + \sum\limits_{n\in 2\mathbb{Z}} a_n(r_p) \cos(n\theta).
\end{align}

In Appendix \ref{app:1}, we demonstrate that the only possible angular dependence when density and ellipticity fields are Gaussian is the $\cos(2\theta)$-term.  Under these conditions, which are met in the LA model, and applying eq.~\eqref{v}, we find:
\begin{align}
\label{t}
w_{g}(r_p,\theta) = w_{g}(r_p) +2 \tilde{w}_{g +}(r_p)\cos(2\theta).
\end{align}
Thus, $w_{g}(r_p,\theta)$ contains the same information as $w_g(r_p)$ and $w_{g+}(r_p)$ in the LA model.

\begin{figure}
\begin{center}
\resizebox{\hsize}{!}{
\includegraphics{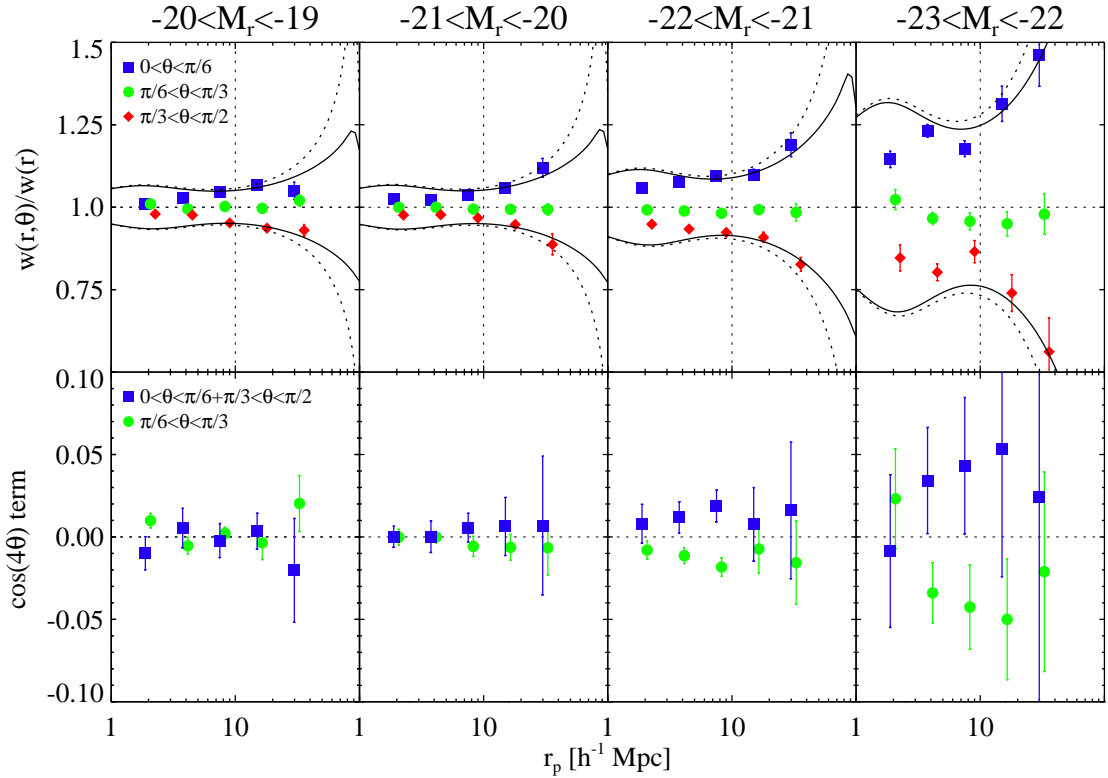}}
\end{center}
\caption{{\it Top panel:} The best-fit LA model prediction for $w_g(r_p,\theta)/w_g(r_p)$ compared with the measurements of \cite{faltenbacher09} for 4 luminosity bins.  The Halofit $P_{\delta}(k)$ is used.  Solid curves include redshift-space distortions, while dashed curves do not.  The measurements are divided into 3 angular bins.  {\it Bottom panel:} Deviations from the prediction of $\cos(2\theta)$ angular dependence. See section 4.2 for more information.  For clarity, small horizontal offsets have been introduced to circle and diamond points.}
\label{fig:wrtheta}
\end{figure}

Figure \ref{fig:wrtheta} compares the LA model prediction of the alignment correlation function with the SDSS main sample measurements of \cite{faltenbacher09}.  These measurements divide galaxies by luminosity and type.  Errors represent the variance between 10 realizations in which galaxy orientations have been randomly shuffled, as described in \cite{faltenbacher09}. We consider red galaxies in the following four magnitude bins: $\{-20<M_r<-19; -21<M_r<-20; -22<M_r<-21; -23<M_r<-22\}$.  Consistent with the measurement, a projection length of $\Pi_{\rm max}=40 h^{-1}$Mpc is used to calculate the LA model prediction.  Galaxy bias for each luminosity bin is calculated by fitting to the observed correlation function (above 10 \hMpc) and assuming that the total sample has $b_g=1.07 \pm 0.01$ \cite{zehavi10}. The best-fit values of LA strength, in order of increasing luminosity, are $\tilde{C}_1 \rho_{\rm crit} = \{0.16 \pm 0.01; 0.17 \pm 0.02; 0.36 \pm 0.03; 1.55 \pm 0.16\}$.
We speculate on what the trend of stronger alignment for more luminous objects may imply about the dynamics of galaxy alignment in section \ref{sec:summ}. The solid curves in figure~\ref{fig:wrtheta} include redshift-space distortions, while the dashed curves do not.  Redshift-space distortions alter the ratio $w_{g}(r_p,\theta)/w_{g}(r_p)$ by over 50\% on large scales.  The galaxy correlation function is more enhanced than the angular term because it involves an integral over $J_0(\kappa r_p)$ rather than $J_2(\kappa r_p)$ and thus receives greater contributions where enhancement is large (see appendix \ref{app:4}).

\subsubsection{Weighting by galaxy ellipticity}

Recent work \cite{okumura09b} has detected a small correlation between $\theta$ and $\gamma_0$ that affects the measured ratio between ellipticity-weighted and unweighted GI statistics on all scales.  In the absence of such a correlation, the ratio between ellipticity-weighted and unweighted statistics is simply the sample mean ellipticity $\langle \gamma_0\rangle$.  This correlation can arise, for instance, if galaxies pointing towards over-dense regions have larger average ellipticities.  This ratio depends on correlations of both $\gamma_0$ and $\delta_g$ with $\theta$.
For Gaussian density and ellipticity fields, we find that the effect \nolinebreak of this correlation is to modulate the ratio at all separations by a factor $4/\pi$ (see Appendix \ref{app:1}):
\begin{align}
\label{weighted to unweighted}
\frac{w^{LA}_{g+}}{\tilde{w}^{LA}_{g+}}= \frac{C_1}{\tilde{C_1}} = \frac{4}{\pi}\langle \gamma_0\rangle = \frac{2\gamma_{{\rm rms}}}{\sqrt{\pi}},
\end{align}
where $\gamma_{\rm rms} \equiv \sqrt{\left\langle \gamma_0^2\right\rangle}$.
The $\approx 15 \pm 10\%$ excess of this ratio found at all separations in \cite{okumura09b} is roughly consistent with the $4/\pi -1 = 27\%$ excess predicted by the LA model.  However, alignment stochasticity may be able to suppress this ratio (section \ref{sec:stoc}).

\subsubsection{Higher-order angular dependence}
Detection of $\cos(n\theta)$ angular dependence in $w_{g}(r_p,\theta)$ for $n \ge 4$ at a given scale would indicate contributions from nonlinearities (which introduce non-Gaussianity in the density field) or other alignment mechanisms.

Detecting such dependence is challenging, further complicated by the fact that small-scale alignment processes suppress these terms (see section \ref{sec:stoc}).  We can fit observed or simulated data to a model which includes a higher-$n$ angular term:
\begin{align}
w(r_p,\theta) = w(r_p) +  2\tilde{w}_{g+}(r_p)\cos(2\theta) + w_4(r_p)\cos(4\theta).
\end{align}
In the case of the binned alignment correlation function presented in \cite{faltenbacher09}, the low angular resolution makes a reliable fit challenging.  Nevertheless, one can look for a potential $\cos(4\theta)$ term by comparing observations with the LA prediction.  In the absence of higher-$n$ terms,
\begin{align}
2w_g(r_p,\pi/6<\theta<\pi/3) = w_g(r_p,0<\theta<\pi/6)+w_g(r_p,\pi/3<\theta<\pi/2) = 2w_g(r_p).
\end{align}
In the bottom panel of figure~\ref{fig:wrtheta}, we show that measured deviations from this prediction are largely consistent with zero, although the errors are large.  Future observations with improved precision and better resolution in alignment angle should allow for detection of (or constraints on) higher-$n$ terms arising from processes outside the LA model.  This analysis can also be applied to halo orientations from simulations.

\section{Additional contributions to intrinsic ellipticity}
\label{sec:stoc}
The LA model as described in section~\ref{sec:model} assumes that tidal alignment occurs down to some minimum scale, determined by the smoothing filter $S$, and processes occurring on smaller scales are ignored. We now consider extending the model by including alignment stochasticity due to the small-scale processes.  For example, this stochasticity could arise from the small-scale tidal field or internal processes within the galaxy that generate ellipticity.  Such stochasticity need not be small and will result in galaxy ellipticities that deviate from the predictions of any large-scale model in an uncorrelated fashion.  The presence of such stochasticity is physically motivated, and \cite{faltenbacher09,okumura09a} have shown that misalignment between halo and galaxy ellipticities is necessary to match the observed alignment amplitude.

We first consider a model in which stochasticity results in a Gaussian-distributed misalignment angle $\theta_m$ with width $\sigma$.  This model has been previously used \cite{okumura09a,okumura09b} to introduce misalignment between the orientations of simulated halos and the galaxies that populate these halos.  The full misalignment between galaxies and the background tidal field is the combination of misalignments between (1) the tidal field and halo and (2) halo and galaxy. Different physical processes may contribute to each.  We define the probability distribution of the misalignment angle, $\theta_m$, with width $\sigma$:
\begin{align}
f(\theta_m)d\theta_m = \frac{1}{\sqrt{2\pi}\sigma}\exp\left[ -\frac{1}{2}\left( \frac{\theta_m}{\sigma}\right)^2\right]d\theta_m.
\end{align}

We now calculate the effect of this misalignment on the alignment statistics.  The true alignment correlation function is the convolution of the original $w_g(r_p,\theta)$ and the misalignment distribution: $w^{\rm true}_g(r_p,\theta)=w_g(r_p,\theta)\ast f(\theta)$.  This convolution provides the effect of stochasticity on $w_{g+}$, which is simply the dipole moment of $w_g(r_p,\theta)$.  Angular dependence in $w_g(r_p,\theta)$ appears as $\cos(n\theta)$ terms, so the relevant convolution is
\begin{align}
\label{misalign conv}
\int d\theta_m\cos(n(\theta-\theta_m))\frac{1}{\sqrt{2\pi}\sigma}\exp\left[ -\frac{1}{2}\left( \frac{\theta_m}{\sigma}\right)^2\right] = \cos(n\theta)\exp\left[ -\frac{1}{2}n^2\sigma^2\right].
\end{align}

For each $\cos(n\theta)$ term, this model leads to a stochastic suppression factor of $S_{n,{\rm stoc}} \equiv \exp\left[ -\frac{1}{2}n^2\sigma^2\right]$.  Stochastic angular misalignment suppresses the IA signal and does so particularly strongly for higher-order effects (i.e. for larger values of $n$).  Physically, this exponential suppression comes from periodicity under rotation: a stochastic rotation of $\sim \pi/n$ erases the signal.  Since $w_{g+}$ comes from the $n=2$ term of the alignment correlation function, stochasticity provides a suppression of $S_{2,{\rm stoc}} = \exp\left[ -2\sigma^2\right]$.  Similarly, $w_{++}$ and $w_{\times \times}$ come from the n=2 term of an analogous alignment correlation function with two independent angles which both contribute in the convolution.  Thus, they will be suppressed by $S_{2,{\rm stoc}}^2=\exp\left[ -4\sigma^2\right]$.  In the previous sections, the effects of stochasticity are included in the values of $C_1$ determined by fitting to measurement.   The magnitude of $C_1$ thus reflects both the degree to which galaxies align with the tidal field and the amount which they are stretched along it.

The analysis of \cite{okumura09a} measured the misalignment between halos in an $N$-body simulation and the resident galaxies required to match the observed alignment amplitude, finding $\sigma \approx 35^{\circ}$ = 0.61 rad.  This value ignores misalignment between the background tidal field and the halos and thus provides a lower limit to the total misalignment of galaxies.  Using this value, $S_{2,{\rm stoc}}\approx0.5$, and the suppression of an $n=4$ term is $\approx$10 times greater.

The angular-misalignment model discussed above does not include any stochastic contribution to the magnitude of ellipticity and is thus more applicable to the unweighted statistics, which do not depend on this magnitude.  To include more general stochastic effects, we consider a second model that assumes a Gaussian distributed scatter in $\gamma_+$ and $\gamma_{\times}$.  For a distribution width of $\Delta_{\gamma}$, this yields stochastic contributions of $\gamma_{+ s}$ and $\gamma_{\times s}$ with
\begin{align}
f(\gamma_{+ s},\gamma_{\times s})d\gamma_{+ s} d\gamma_{\times s} = \frac{1}{\pi\Delta_{\gamma}^2}\exp\left[ -\left( \frac{\gamma_{+ s}^2+\gamma_{\times s}^2}{\Delta_{\gamma}^2}\right)\right]d\gamma_{+ s} d\gamma_{\times s}.
\end{align}

Note that this model is equivalent to adding Gaussian scatter in the perpendicular components of observed galaxy shape.  The effect of stochasticity in this model is to increase the total value of galaxy ellipticity: $\gamma_{\rm rms, total}^2= \gamma_{\rm rms, LA}^2 + \Delta_{\gamma}^2$.  Since it adds only a non-correlated component to ellipticity, this model does not affect the magnitude of $w_{g+}$ relative to the LA model with no stochasticity: $\langle\delta \, (\gamma_+^{\rm LA}+\gamma_+^{\rm stoc})\rangle = \langle\delta \gamma_+^{\rm LA}\rangle$.  However, by increasing $\gamma_{\rm rms, total}$, it suppresses $\tilde{w}_{g+}$ and thus the $\cos(2\theta)$-term of $w_g(r_p,\theta)$.  The stochastic contributions to $\gamma_+$ and $\gamma_{\times}$ result in both a change in the observed ellipticity magnitude and alignment angle.  The suppression  of $\tilde{w}_{g+}$ can be seen as coming from the angular misalignment, which is qualitatively similar to the angular misalignment model discussed above.  Both models thus result in suppressed angular information, particularly for higher-$n$ terms.

\begin{figure}[t!]
\begin{center}
\includegraphics[width=8.8cm]{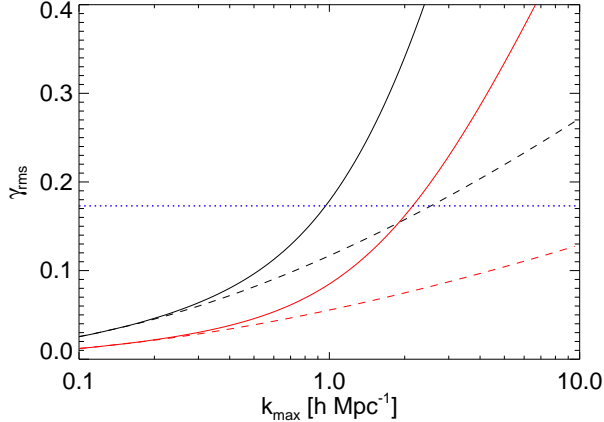}
\end{center}
\caption{The LA model prediction for $\gamma_{\rm rms}$ plotted as a function of the cut-off value $k_{\rm max}$.  Solid lines are calculated using the Halofit $P_{\delta}(k)$, while the dashed lines are calculated with the linear $P_{\delta}(k)$.  The prediction is normalized assuming the best-fit LRG value for $C_1$.  Black lines assume a level of stochasticity consistent with \cite{okumura09a}, while red lines assume that no stochasticity is present.  For reference, the observed LRG value of $\gamma_{\rm rms}\approx 0.17$ is plotted (horizontal blue dotted line).}
\label{fig:gamma_rms}
\end{figure}

In addition to affecting the alignment correlation statistics, astrophysical stochasticity alters the observed value of \grms.  In principle, one can measure the magnitude of stochasticity by comparing measurements of \grms\ with the LA theory prediction in the absence of stochasticity: $\gamma^2_{\rm rms, LA} = 2\xi^{\rm LA}_{++}(r_p=0)$.  The true \grms\ will exceed the stochasticity-free prediction.  Although stochasticity in the angular-misalignment model makes no contribution to \grms, it suppresses the measured value of $C_1$ by $S_{2,{\rm stoc}}$, leading to an inferred \grms\ less than the true value.  However, it is challenging to draw conclusions from this comparison because the integral for $\xi^{\rm LA}_{++}(0)$ is highly dependent on the smallest scales for which the LA model is assumed to be valid (i.e. the choice of smoothing filter).  This dependence is illustrated in figure \ref{fig:gamma_rms}.  For instance, smoothing on a typical halo scale corresponds to a maximum wavevector $k_{\rm max} \sim 2\pi/1$ \hMpcinv, while smoothing on nonlinear scales corresponds to $k_{\rm max} \sim 2\pi/10$ \hMpcinv.  Due to these complications, we have chosen to leave $\gamma_{\rm rms}$ as a parameter to be measured independently.  The value of $C_1$ can then be determined from the alignment statistics rather than from $\gamma_{\rm rms}$, as has been sometimes done in the absence of alignment measurements (e.g.~\cite{catelan01,hirata04}).  Nevertheless, as seen in figure \ref{fig:gamma_rms}, alignment with the large-scale tidal field can account for a large fraction of the total observed LRG ellipticity. Note that for $k_{\rm max}\gtrsim 1-10$ \hMpcinv, the LA model will start to break down and the predictions shown here will be of limited validity.

Finally, we consider whether stochasticity suppresses the ratio $w_{g+}/\tilde{w}_{g+}$, discussed in \S3.4.  In general, stochasticity can suppress this ratio, which depends on correlations of both $\gamma_0$ and $\delta_g$ with $\theta$.  However, in the two models considered in this section, the ratio is unaltered because any decrease in the correlation between $\gamma_0$ and $\theta$ is exactly offset by a decrease in the correlation between $\delta_g$ and $\theta$.

\section{Summary and conclusions}
\label{sec:summ}
This work examined intrinsic galaxy alignment in the context of the tidal alignment model.  Using both II and GI correlation statistics in configuration space, we showed that the model is consistent with measurements on scales above $\gtrsim10~$\hMpc.  Including nonlinear corrections in $P_{\delta}(k)$ improves the fit at smaller separations, although we note that this is not a rigorous nonlinear model.  We measured best-fit values of the model strength parameter $C_1$ using several statistics and two data sets, finding results that are consistent both internally and with previous measurements.  We examined IA statistics both unweighted and weighted by ellipticity magnitude, $\gamma_0$.  The ratio of these statistics (e.g.~$w_{g+}/\tilde{w}_{g+}$) depends not only on the mean value of ellipticity for the observed galaxy sample, $\langle \gamma_0\rangle$, but also on correlations between both $\gamma_0$ and $\delta_g$ and the position angle, $\theta$.  The LA model makes predictions for these correlations which are qualitatively similar to what was observed in \cite{okumura09b}.

The LA model predicts zero $B$-mode auto-correlation signal at lowest order.  We found that LRG measurements are consistent with this prediction.  More precise measurements of the $B$-mode signal should provide a valuable way to determine the importance of higher-order terms in $\delta$ in the LA model, other alignment mechanisms, and systematic errors.

The recently proposed alignment correlation function $w_g(r_p,\theta)$ is of particular interest as its angular dependence can contain additional information on galaxy alignment and large-scale structure.  However, we showed that for Gaussian density and ellipticity fields, a condition satisfied in the LA model, $w_g(r_p,\theta)$ has angular dependence $\cos(2\theta)$ and is completely described by $w_g(r_p)$ and $w_{g+}(r_p)$.  This angular dependence is consistent with the recent measurement of \cite{faltenbacher09}.  We found that departures from these conditions allow additional angular terms to contribute, although they are suppressed by alignment stochasticity.  The next order angular term is suppressed by an order of magnitude for the level of misalignment that has been measured \cite{okumura09a,okumura09b}.

Finally, we considered how to include stochastic contributions to a model of intrinsic alignment.  Astrophysical sources of stochasticity can affect both the magnitude and direction of galaxy ellipticity, and previous measurements \cite{okumura09a,okumura09b,faltenbacher09} indicate that such stochasticity is needed to account for differences between the strength of alignment measured in simulated halos as opposed to observed galaxies.  This stochasticity is compounded by uncertainties in measuring $\langle \gamma_0\rangle$: correcting for the PSF can alter the observed value by $\approx$\;10\,-\,30\%.  Due to these complications, it is not recommended to normalize any IA model using local statistics such as ellipticity mean or variance as some studies have done.  Nevertheless, the LA model is able to account for a significant fraction of observed LRG ellipticity.  We considered two models for stochasticity, angular misalignment and ellipticity scatter, and found that these models have similar observational signatures and are consistent with current measurements of intrinsic alignment amplitude.  Since much of this stochasticity should come from physical processes such as galaxy mergers and gas dynamics on nonlinear scales, it is likely that the stochastic suppression will be a function of time, and thus care must be taken when attempting to separate the intrinsic alignment and weak lensing signals based on their redshift dependence (e.g.~\cite{catelan01}).

Previous work (e.g.~\cite{faltenbacher09,joachimi10}) has found a significant luminosity dependence in the strength of galaxy alignment.  The tendency of stochasticity to vary with time will affect the relative amplitudes of alignment correlation statistics between galaxy samples that formed at different redshifts.  One possible explanation for the trend of alignment increasing with galaxy luminosity is that more luminous objects have formed more recently and have had less time to misalign from the tidal axis along which they formed.  This picture is consistent with a scenario in which halo ellipticities are determined primarily during times of collapse and accretion.  For instance, one plausible mechanism for forming triaxial halos aligned with the tidal axis is the radial orbit instability (ROI; see, e.g., \cite{macmillan06,carpintero95}).  The ROI occurs when particles with radial orbits are torqued by a small asymmetry in an otherwise spherical system, causing them to align with and enhance the asymmetry.  The ROI is potentially responsible for the overall triaxial shape as well as the nearly universal radial density profile of halos \cite{macmillan06,carpintero95,bellovary08}.  We speculate that the ROI (or some other mechanism) acts during the accretion of matter to align the halo and the galaxy it contains with the tidal field.  Such a mechanism may allow the accretion of a small fraction of total halo mass to affect the shape and orientation of the entire halo.

However, this alignment mechanism is likely a transient phenomenon.  After accretion stops, the shape and orientation of the halo should evolve on roughly a dynamical timescale.  Because dynamical time decreases with density, the outer region of a halo will have significantly longer dynamical times than the central region, where the galaxy resides.  Thus, the inner region (and associated galaxy) will acquire a misalignment from the tidal field more quickly than the halo as a whole. The process of misalignment has been studied in simulations by \cite{pereira10}, who find that halos orbiting in a cluster environment can experience significant internal twisting leading to appreciable misalignment between dark matter and a central stellar component.

This connection of matter accretion to galaxy orientation yields two predictions.  First, objects that have more recently accreted matter will be more strongly aligned with the tidal field (corresponding to a larger value of $C_1$ in the LA model).  This trend is consistent with the observed luminosity dependence of the IA signal.  Second, because of the gradient in dynamical time, stochastic evolution after the shut-off of the alignment mechanism, via gravitational relaxation or interactions with other systems, will result in misalignment between the central regions of halos (or equivalently the galaxies they contain) and the outer halo regions, which will remain more closely oriented with the tidal field.  Such misalignment is observed in simulations \cite{faltenbacher09} with an amplitude consistent with that required to match alignment statistics calculated from simulations with those observed from real galaxies \cite{okumura09a,okumura09b}.  Using a semi-analytic galaxy catalog approach, \cite{faltenbacher09} find that more luminous galaxies (assumed to have the same orientation as the central halo component) are more closely aligned with their parent halos.  However, the trend they measure is too weak to account for all of the luminosity dependence in the alignment correlation function.  Thus, halos containing more luminous objects may also exhibit stronger alignment with the tidal field, and galaxies may be less aligned with the central halo region than assumed.

It would be interesting to more closely examine possible mechanisms for establishing intrinsic alignment. The interplay between accretion and other dynamical evolution may impart a redshift dependence on the strength of alignment. While accretion was generally more rapid at higher redshift, at later times the effects of additional accretion could increase alignment, or other evolution processes could become significant. Accurate characterization of this redshift dependence would improve our understanding of the importance of different physical processes in galaxy formation and evolution (although, see \cite{joachimi10} who find no significant redshift dependence for LRGs out to $z\approx 0.7$).  Further measurements and theoretical work in this area are of particular relevance since the sources used in lensing studies are often at higher redshifts than those used to measure IA.  Future work with both $N$-body and baryon simulations can probe the dynamical aspects of galaxy alignment.  Detailed observations could provide a direct probe of misalignment between galaxies and the halos in which they reside.

Encouraging progress has been made in recent years towards understanding the intrinsic alignment of galaxies and quantifying its effects.  Much work remains to be done as we prepare for the next generation of weak lensing experiments. The issues discussed here, combined with the improved precision of future measurements, will provide methods to better describe and account for the intrinsic alignment of galaxies.

\acknowledgments

We thank Teppei Okumura for several useful discussions and Rachel Mandelbaum for comments on the manuscript.  We also thank Cheng Li and Andreas Faltenbacher for sharing their measurements and Michael Schneider for helpful discussion. Finally, we are grateful for comments from an anonymous referee. This work is supported by the DOE, the Swiss National Foundation under contract 200021-116696/1, WCU grant R32-2009-000-10130-0, and an NSF Graduate Research Fellowship.  M.M. is supported by a NASA Einstein Fellowship.

\appendix
\section{Calculating $w(r_p,\theta)$ in the linear alignment model}
\label{app:1}
Consider a set of $N$ Gaussian fields $\{q_i\}$, which are assembled into a vector $\boldsymbol{q}$, and some function of these fields $F(q_1,...,q_N)$.  An ensemble average of $F$ is given by the integral:
\begin{align}
\label{a1}
\langle F\rangle = \int dq_1...dq_N \frac{F(q_1,...,q_N)}{(2\pi)^{N/2}(detC)^{1/2}}e^{-\frac{1}{2}\vec{q}^TC^{-1}\vec{q}}
\end{align}
where $C_{ij} \equiv \langle q_i q_j\rangle$ is the relevant covariance matrix.

We consider a reference galaxy at position $\boldsymbol{x_1}$ with shear $\gamma_0$ and a second galaxy at point $\boldsymbol{x_2}$.  The second galaxy has position angle $\theta$, as measured from the major axis of the reference galaxy.  Measured from the separation axis, the reference galaxy has $\gamma_+ = \gamma_0\cos(2\theta)$.  We are interested in calculating the 3-dimensional alignment correlation function, $\xi_g(\boldsymbol{r},\theta)$.  We start with the definition of the standard correlation function, $\xi_g(\boldsymbol{r})$:
\begin{align}
\label{a4}
\langle n_1n_2\rangle \equiv \bar{n}^2[1+\xi_g(\boldsymbol{r})],
\end{align}
where $n_{1} = \bar{n}[1+\delta_{1}]$ and from the $\theta$-dependence, $n_{2} = \bar{n}[1+\delta_{2}(\theta)]$.  Thus,
\begin{align}
\label{a5}
\langle n_1n_2|\theta\rangle = \bar{n}^2[1 + \langle \delta_2|\theta\rangle + \langle \delta_1\delta_2|\theta\rangle].
\end{align}
Comparing with \eqref{a4}, we have:
\begin{align}
\label{a6}
\xi_g(\boldsymbol{r},\theta) = \langle \delta_2|\theta\rangle + \langle \delta_1\delta_2|\theta\rangle.
\end{align}

We apply eq. \eqref{a1} with $F=\delta_2 + \delta_1 \delta_2$ and four Gaussian fields: $\{\delta_1, \delta_2, \gamma_{+}, \gamma_{\times}\}$, where $\gamma_{(+,\times)}$ is evaluated for the reference galaxy at $\boldsymbol{x_1}$.  Isotropy gives $\langle \gamma_+\gamma_+\rangle = \langle \gamma_{\times}\gamma_{\times}\rangle$, $\langle \delta_1\gamma_{(+,\times)}\rangle=0$, and parity conservation requires $\langle \gamma_+\gamma_{\times}\rangle  = \langle \delta_2\gamma_{\times}\rangle = 0$.  These symmetries greatly simplify the covariance matrix.  Finally, we switch integration variables using $d\gamma_+d\gamma_{\times} = d\gamma_0d\theta \,{\rm det}(J)$, where $J$ is the Jacobian and ${\rm det}(J) = 2\gamma_0$.  Integrating over all variables except for $\theta$, the result is
\begin{align}
\label{a7}
\xi_g(\boldsymbol{r},\theta) = \xi_g(\boldsymbol{r}) +\left(\frac{\pi}{2\langle\gamma_+ \gamma_+\rangle}\right)^{1/2}\cos(2\theta)\langle \delta_2\gamma_+\rangle_{\boldsymbol{r}},
\end{align}
where we have multiplied by an overall factor of $\pi$ to match the normalization convention adopted in the paper.  Performing the projection yields
\begin{align}
\label{a8}
w_g(r_p,\theta) = w_g(r_p) +\left(\frac{\pi}{2\langle\gamma_+ \gamma_+\rangle}\right)^{1/2}\cos(2\theta)w_{g+}(r_p).
\end{align}
We thus see that Gaussian density and ellipticity fields allow only $\cos(2\theta)$ angular dependence in $w_g(r_p,\theta)$.  From eqs. \eqref{v} and \eqref{a8}, we also find the following result:
\begin{align}
\label{a9}
\frac{w_{g+}}{\tilde{w}_{g+}} = \frac{2\gamma_{\rm rms}}{\sqrt{\pi}},
\end{align}
where we have used $2\langle\gamma_+ \gamma_+\rangle=\gamma^2_{\rm rms}$.  This ratio is the effective average of $\gamma_0$, where the ellipticity of each galaxy is weighted by how strongly it contributes to the statistic.  If there were no correlation between $\gamma_0$ and $\theta$, the ratio would simply be $\langle\gamma_0\rangle$ for the entire galaxy sample.  It is this hypothesis that is tested in \cite{okumura09b}, where they detect a $\sim 15\%$ modulation effect due to a correlation between $\gamma_0$ and $\theta$.  Note that the results derived here (eqs. \eqref{a7}-\eqref{a9}) hold for any scenario in which the density and ellipticity fields are Gaussian, such as the LA model.

We can employ the same formalism to more closely examine this correlation in the LA model.  If we don't perform the $\gamma_0$ integral, we find:
\begin{align}
\label{a10}
w_g(r_p,\theta,\gamma_0) = \frac{\gamma_0}{\pi \langle\gamma_+\gamma_+\rangle^2}\exp\left(-\frac{\gamma_0^2}{2\langle\gamma_+\gamma_+\rangle}\right)\left[\langle\gamma_+^2\rangle w_g(r_p) +\gamma_0\cos(2\theta)w_{g +}(r_p)\right].
\end{align}
We can define an ellipticity-weighted alignment correlation function $w_g^{\gamma}(r_p,\theta)$. Without the $\gamma_0$ integral, this statistic is $w_g^{\gamma}(r_p,\theta,\gamma_0) = \gamma_0\, w_g(r_p,\theta,\gamma_0)$.

The isotropic term gives the standard correlation function $w_g(r_p)$, while the angle-dependent term is related to $w_{g +}$.  These terms have different $\gamma_0$ scalings because $\delta_2$ and $\delta_1 \delta_2$ pick out different terms in the exponent of the likelihood integral, eq.~\eqref{a1}.  The ratios between unweighted and weighted statistics will thus be different for isotropic and angle-dependent terms.  Performing the integral over $\gamma_0$ yields
\begin{align}
\label{a11}
\frac{w_{g+}(r_p)}{\tilde{w}_{g+}(r_p)}=\frac{4}{\pi}\frac{w_g^{\gamma}(r_p)}{w_g(r_p)}=\frac{2\gamma_{\rm rms}}{\sqrt{\pi}}.
\end{align}

When we only consider counts of galaxy pairs, as with $w_g(r_p)$, all galaxies contribute equally.  When pair counts are multiplied by alignment angle, as with $w_{g +}$, galaxies that are more closely aligned with separation vectors are more heavily weighted.  A correlation between $\gamma_0$ and $\theta$ leads to a different effective average value for $\gamma_0$.  For the LA model, such a correlation is present: more closely aligned galaxies have larger average values of $\gamma_0$.

\section{Bessel function identities for E and B modes}
\label{app:3}
Inserting the LA model results into eq. \eqref{m}, we find:
\begin{align}
w'(x) = &\frac{1}{\pi^2}\left(\frac{C_1\rho_{m,0}}{D(z)}\right)^2\int_0^{\infty} d\kappa dk_z \frac{\kappa^5}{k^4 k_z}P_{\delta}(k,z)\sin(k_z \Pi_{\rm max})\\
\notag
&\times\left[ J_4(\kappa x) + 4\int_{x}^{\infty}dx'\frac{J_4(\kappa x')}{x'} - 12x^2\int_x^{\infty}dx'\frac{J_4(\kappa x')}{x'^3}\right].
\end{align}
Applying the recursion relation $J_n(y) = \frac{y}{2n}\left[ J_{n-1}(y) + J_{n+1}(y)\right]$ yields:
\begin{align}
J_0(y) = J_4(y)+\frac{8}{y}\left( \frac{1}{y}J_2(y) - J_3(y) \right).
\end{align}
We then use following integral identities:
\begin{align}
\label{bessel}
\int_x^{\infty}dx'\frac{J_4(\kappa x')}{x'} &= \frac{1}{(\kappa x)^2}\left[ 2J_2(\kappa x) + \kappa x J_3(\kappa x)\right],\\
\notag
\int_x^{\infty}dx'\frac{J_4(\kappa x')}{x'^3} &= \frac{J_3(\kappa x)}{\kappa x^3},
\end{align}
and find
\begin{align}
J_4(\kappa x) + 4\int_x^{\infty}dx'\frac{J_4(\kappa x')}{x'} - 12x^2\int_x^{\infty}dx'\frac{J_4(\kappa x')}{x'^3} = J_0(\kappa x).
\end{align}
Thus, $w'(x) = w_+(x)$ in the LA model.

\section{Including redshift-space distortions}
\label{app:4}
When measured redshifts are used to determine distances to galaxies, peculiar velocities introduce errors in the line-of-sight position \cite{kaiser87}.  For the LA model, we are concerned with large scales, where coherent infall causes an apparent squashing of structure along the line-of-sight, leading to a spurious enhancement of clustering power.  As first described in \cite{kaiser87}, we can relate the power spectrum in real space, $P_r(k)$,  with one measured in redshift space, $P_s(k)$:
\begin{align}
\label{rss}
P_s(k) = P_r(k)\left[1+\beta \mu^2\right]^2
\end{align}
where $\mu = \cos(\theta_k)$ for angle $\theta_k$ between $\boldsymbol{k}$ and the line-of-sight; $\beta = f(a)/b_g$ for galaxy bias $b_g$ and $f(a)\equiv {\rm d ln}D/{\rm d ln}a$, the logarithmic growth rate of fluctuations.  For this work, we apply the fitting formula $f(a)=\Omega_m(a)^{0.6}$ and values for $b_g$ measured using the amplitude of the correlation functions of the different galaxy samples.  We use $\beta \approx 0.3$ for the LRG sample and $\beta \approx 0.4$ for the main sample, consistent with previous measurements \cite{tegmark06}.  Although we consider projected statistics, the effects of redshift-space distortions remain non-negligible because of the finite projection length (see, e.g., \cite{baldauf10}).


\begin{thebibliography}{10}

\bibitem{blandford91}
R.~D. {Blandford}, A.~B. {Saust}, T.~G. {Brainerd}, and J.~V. {Villumsen}, {\it
  {The distortion of distant galaxy images by large-scale structure}},  {\em
  \mnras} {\bf 251} (Aug., 1991) 600--627.

\bibitem{miralda91}
J.~{Miralda-Escude}, {\it {The correlation function of galaxy ellipticities
  produced by gravitational lensing}},  {\em \apj} {\bf 380} (Oct., 1991) 1--8.

\bibitem{kaiser92}
N.~{Kaiser}, {\it {Weak gravitational lensing of distant galaxies}},  {\em
  \apj} {\bf 388} (Apr., 1992) 272--286.

\bibitem{refregier03}
A.~{Refregier}, {\it {Weak Gravitational Lensing by Large-Scale Structure}},
  {\em \araa} {\bf 41} (2003) 645--668,
  [\href{http://xxx.lanl.gov/abs/astro-ph/0307212}{{\tt astro-ph/0307212}}].

\bibitem{bacon00}
D.~J. {Bacon}, A.~R. {Refregier}, and R.~S. {Ellis}, {\it {Detection of weak
  gravitational lensing by large-scale structure}},  {\em \mnras} {\bf 318}
  (Oct., 2000) 625--640, [\href{http://xxx.lanl.gov/abs/astro-ph/0003008}{{\tt
  astro-ph/0003008}}].

\bibitem{kaiser00}
N.~{Kaiser}, G.~{Wilson}, and G.~A. {Luppino}, {\it {Large-Scale Cosmic Shear
  Measurements}},  {\em \rm ArXiv e-prints} (Mar., 2000)
  [\href{http://xxx.lanl.gov/abs/astro-ph/0003338}{{\tt astro-ph/0003338}}].

\bibitem{vanWaerbeke00}
L.~{Van Waerbeke}, Y.~{Mellier}, T.~{Erben}, J.~C. {Cuillandre},
  F.~{Bernardeau}, R.~{Maoli}, E.~{Bertin}, H.~J. {McCracken}, O.~{Le
  F{\`e}vre}, B.~{Fort}, M.~{Dantel-Fort}, B.~{Jain}, and P.~{Schneider}, {\it
  {Detection of correlated galaxy ellipticities from CFHT data: first evidence
  for gravitational lensing by large-scale structures}},  {\em \aap} {\bf 358}
  (June, 2000) 30--44, [\href{http://xxx.lanl.gov/abs/astro-ph/0002500}{{\tt
  astro-ph/0002500}}].

\bibitem{wittman00}
D.~M. {Wittman}, J.~A. {Tyson}, D.~{Kirkman}, I.~{Dell'Antonio}, and
  G.~{Bernstein}, {\it {Detection of weak gravitational lensing distortions of
  distant galaxies by cosmic dark matter at large scales}},  {\em \nat} {\bf
  405} (May, 2000) 143--148,
  [\href{http://xxx.lanl.gov/abs/astro-ph/0003014}{{\tt astro-ph/0003014}}].

\bibitem{scoville07}
N.~{Scoville} {\em et.~al.}, {\it {The Cosmic Evolution Survey (COSMOS):
  Overview}},  {\em \apjs} {\bf 172} (Sept., 2007) 1--8,
  [\href{http://xxx.lanl.gov/abs/astro-ph/0612305}{{\tt astro-ph/0612305}}].

\bibitem{reyes10}
R.~{Reyes}, R.~{Mandelbaum}, U.~{Seljak}, T.~{Baldauf}, J.~E. {Gunn},
  L.~{Lombriser}, and R.~E. {Smith}, {\it {Confirmation of general relativity
  on large scales from weak lensing and galaxy velocities}},  {\em \nat} {\bf
  464} (Mar., 2010) 256--258, [\href{http://xxx.lanl.gov/abs/1003.2185}{{\tt
  arXiv:1003.2185}}].

\bibitem{daniel10}
S.~F. {Daniel}, E.~V. {Linder}, T.~L. {Smith}, R.~R. {Caldwell}, A.~{Cooray},
  A.~{Leauthaud}, and L.~{Lombriser}, {\it {Testing general relativity with
  current cosmological data}},  {\em \prd} {\bf 81} (June, 2010) 123508--+,
  [\href{http://xxx.lanl.gov/abs/1002.1962}{{\tt arXiv:1002.1962}}].

\bibitem{thomas09}
S.~A. {Thomas}, F.~B. {Abdalla}, and J.~{Weller}, {\it {Constraining modified
  gravity and growth with weak lensing}},  {\em \mnras} {\bf 395} (May, 2009)
  197--209, [\href{http://xxx.lanl.gov/abs/0810.4863}{{\tt arXiv:0810.4863}}].

\bibitem{lombriser10}
L.~{Lombriser}, A.~{Slosar}, U.~{Seljak}, and W.~{Hu}, {\it {Constraints on
  f(R) gravity from probing the large-scale structure}},  {\em \rm ArXiv
  e-prints} (Mar., 2010) [\href{http://xxx.lanl.gov/abs/1003.3009}{{\tt
  arXiv:1003.3009}}].

\bibitem{albrecht06}
A.~{Albrecht}, G.~{Bernstein}, R.~{Cahn}, W.~L. {Freedman}, J.~{Hewitt},
  W.~{Hu}, J.~{Huth}, M.~{Kamionkowski}, E.~W. {Kolb}, L.~{Knox}, J.~C.
  {Mather}, S.~{Staggs}, and N.~B. {Suntzeff}, {\it {Report of the Dark Energy
  Task Force}},  {\em \rm ArXiv e-prints} (Sept., 2006)
  [\href{http://xxx.lanl.gov/abs/astro-ph/0609591}{{\tt astro-ph/0609591}}].

\bibitem{simon07}
P.~{Simon}, M.~{Hetterscheidt}, M.~{Schirmer}, T.~{Erben}, P.~{Schneider},
  C.~{Wolf}, and K.~{Meisenheimer}, {\it {GaBoDS: The Garching-Bonn Deep
  Survey. VI. Probing galaxy bias using weak gravitational lensing}},  {\em
  \aap} {\bf 461} (Jan., 2007) 861--879,
  [\href{http://xxx.lanl.gov/abs/astro-ph/0606622}{{\tt astro-ph/0606622}}].

\bibitem{mandelbaum06a}
R.~{Mandelbaum}, C.~M. {Hirata}, M.~{Ishak}, U.~{Seljak}, and J.~{Brinkmann},
  {\it {Detection of large-scale intrinsic ellipticity-density correlation from
  the Sloan Digital Sky Survey and implications for weak lensing surveys}},
  {\em \mnras} {\bf 367} (Apr., 2006) 611--626,
  [\href{http://xxx.lanl.gov/abs/astro-ph/0509026}{{\tt astro-ph/0509026}}].

\bibitem{hirata04}
C.~M. {Hirata} and U.~{Seljak}, {\it {Intrinsic alignment-lensing interference
  as a contaminant of cosmic shear}},  {\em \prd} {\bf 70} (Sept., 2004)
  063526--+, [\href{http://xxx.lanl.gov/abs/astro-ph/0406275}{{\tt
  astro-ph/0406275}}].

\bibitem{croft00}
R.~A.~C. {Croft} and C.~A. {Metzler}, {\it {Weak-Lensing Surveys and the
  Intrinsic Correlation of Galaxy Ellipticities}},  {\em \apj} {\bf 545} (Dec.,
  2000) 561--571, [\href{http://xxx.lanl.gov/abs/astro-ph/0005384}{{\tt
  astro-ph/0005384}}].

\bibitem{heavens00}
A.~{Heavens}, A.~{Refregier}, and C.~{Heymans}, {\it {Intrinsic correlation of
  galaxy shapes: implications for weak lensing measurements}},  {\em \mnras}
  {\bf 319} (Dec., 2000) 649--656,
  [\href{http://xxx.lanl.gov/abs/astro-ph/0005269}{{\tt astro-ph/0005269}}].

\bibitem{catelan01}
P.~{Catelan}, M.~{Kamionkowski}, and R.~D. {Blandford}, {\it {Intrinsic and
  extrinsic galaxy alignment}},  {\em \mnras} {\bf 320} (Jan., 2001) L7--L13,
  [\href{http://xxx.lanl.gov/abs/astro-ph/0005470}{{\tt astro-ph/0005470}}].

\bibitem{crittenden01}
R.~G. {Crittenden}, P.~{Natarajan}, U.~{Pen}, and T.~{Theuns}, {\it
  {Spin-induced Galaxy Alignments and Their Implications for Weak-Lensing
  Measurements}},  {\em \apj} {\bf 559} (Oct., 2001) 552--571,
  [\href{http://xxx.lanl.gov/abs/astro-ph/0009052}{{\tt astro-ph/0009052}}].

\bibitem{jing02}
Y.~P. {Jing}, {\it {Intrinsic correlation of halo ellipticity and its
  implications for large-scale weak lensing surveys}},  {\em \mnras} {\bf 335}
  (Oct., 2002) L89--L93, [\href{http://xxx.lanl.gov/abs/astro-ph/0206098}{{\tt
  astro-ph/0206098}}].

\bibitem{king02}
L.~{King} and P.~{Schneider}, {\it {Suppressing the contribution of intrinsic
  galaxy alignments to the shear two-point correlation function}},  {\em \aap}
  {\bf 396} (Dec., 2002) 411--418,
  [\href{http://xxx.lanl.gov/abs/astro-ph/0208256}{{\tt astro-ph/0208256}}].

\bibitem{king03}
L.~J. {King} and P.~{Schneider}, {\it {Separating cosmic shear from intrinsic
  galaxy alignments: Correlation function tomography}},  {\em \aap} {\bf 398}
  (Jan., 2003) 23--30, [\href{http://xxx.lanl.gov/abs/astro-ph/0209474}{{\tt
  astro-ph/0209474}}].

\bibitem{heymans03}
C.~{Heymans} and A.~{Heavens}, {\it {Weak gravitational lensing: reducing the
  contamination by intrinsic alignments}},  {\em \mnras} {\bf 339} (Mar., 2003)
  711--720, [\href{http://xxx.lanl.gov/abs/astro-ph/0208220}{{\tt
  astro-ph/0208220}}].

\bibitem{takada04}
M.~{Takada} and M.~{White}, {\it {Tomography of Lensing Cross-Power Spectra}},
  {\em \apjl} {\bf 601} (Jan., 2004) L1--L4,
  [\href{http://xxx.lanl.gov/abs/astro-ph/0311104}{{\tt astro-ph/0311104}}].

\bibitem{hirata07}
C.~M. {Hirata}, R.~{Mandelbaum}, M.~{Ishak}, U.~{Seljak}, R.~{Nichol}, K.~A.
  {Pimbblet}, N.~P. {Ross}, and D.~{Wake}, {\it {Intrinsic galaxy alignments
  from the 2SLAQ and SDSS surveys: luminosity and redshift scalings and
  implications for weak lensing surveys}},  {\em \mnras} {\bf 381} (Nov., 2007)
  1197--1218, [\href{http://xxx.lanl.gov/abs/astro-ph/0701671}{{\tt
  astro-ph/0701671}}].

\bibitem{faltenbacher09}
A.~{Faltenbacher}, C.~{Li}, S.~D.~M. {White}, Y.~{Jing}, {Shu-DeMao}, and
  J.~{Wang}, {\it {Alignment between galaxies and large-scale structure}},
  {\em \rm Research in Astronomy and Astrophysics} {\bf 9} (Jan., 2009) 41--58,
  [\href{http://xxx.lanl.gov/abs/0811.1995}{{\tt arXiv:0811.1995}}].

\bibitem{okumura09a}
T.~{Okumura}, Y.~P. {Jing}, and C.~{Li}, {\it {Intrinsic Ellipticity
  Correlation of SDSS Luminous Red Galaxies and Misalignment with Their Host
  Dark Matter Halos}},  {\em \apj} {\bf 694} (Mar., 2009) 214--221,
  [\href{http://xxx.lanl.gov/abs/0809.3790}{{\tt arXiv:0809.3790}}].

\bibitem{okumura09b}
T.~{Okumura} and Y.~P. {Jing}, {\it {The Gravitational Shear-Intrinsic
  Ellipticity Correlation Functions of Luminous Red Galaxies in Observation and
  in the {$\Lambda$}CDM Model}},  {\em \apjl} {\bf 694} (Mar., 2009) L83--L86,
  [\href{http://xxx.lanl.gov/abs/0812.2935}{{\tt arXiv:0812.2935}}].

\bibitem{schneider10}
M.~D. {Schneider} and S.~{Bridle}, {\it {A halo model for intrinsic alignments
  of galaxy ellipticities}},  {\em \mnras} {\bf 402} (Mar., 2010) 2127--2139,
  [\href{http://xxx.lanl.gov/abs/0903.3870}{{\tt arXiv:0903.3870}}].

\bibitem{joachimi10}
B.~{Joachimi}, R.~{Mandelbaum}, F.~B. {Abdalla}, and S.~L. {Bridle}, {\it
  {Constraints on intrinsic alignment contamination of weak lensing surveys
  using the MegaZ-LRG sample}},  {\em \rm ArXiv e-prints} (Aug., 2010)
  [\href{http://xxx.lanl.gov/abs/1008.3491}{{\tt arXiv:1008.3491}}].

\bibitem{lee08a}
J.~{Lee}, V.~{Springel}, U.~{Pen}, and G.~{Lemson}, {\it {Quantifying the
  cosmic web - I. The large-scale halo ellipticity-ellipticity and
  ellipticity-direction correlations}},  {\em \mnras} {\bf 389} (Sept., 2008)
  1266--1274, [\href{http://xxx.lanl.gov/abs/0709.1106}{{\tt
  arXiv:0709.1106}}].

\bibitem{hui02}
L.~{Hui} and J.~{Zhang}, {\it {Intrinsic/Extrinsic Density-Ellipticity
  Correlations and Galaxy-Galaxy Lensing}},  {\em \rm ArXiv e-prints} (May,
  2002) [\href{http://xxx.lanl.gov/abs/astro-ph/0205512}{{\tt
  astro-ph/0205512}}].

\bibitem{ciotti94}
L.~{Ciotti} and S.~N. {Dutta}, {\it {Alignment and Morphology of Elliptical
  Galaxies - the Influence of the Cluster Tidal Field}},  {\em \mnras} {\bf
  270} (Sept., 1994) 390--400,
  [\href{http://xxx.lanl.gov/abs/astro-ph/9404059}{{\tt astro-ph/9404059}}].

\bibitem{ciotti98}
L.~{Ciotti} and G.~{Giampieri}, {\it {Motion of a rigid body in a tidal field:
  an application to elliptical galaxies in clusters}},  {\em ArXiv Astrophysics
  e-prints} (Jan., 1998) [\href{http://xxx.lanl.gov/abs/astro-ph/9801261}{{\tt
  astro-ph/9801261}}].

\bibitem{pereira08}
M.~J. {Pereira}, G.~L. {Bryan}, and S.~P.~D. {Gill}, {\it {Radial Alignment in
  Simulated Clusters}},  {\em \apj} {\bf 672} (Jan., 2008) 825--833,
  [\href{http://xxx.lanl.gov/abs/0707.1702}{{\tt arXiv:0707.1702}}].

\bibitem{pereira10}
M.~J. {Pereira} and G.~L. {Bryan}, {\it {Tidal Torquing of Elliptical Galaxies
  in Cluster Environments}},  {\em \apj} {\bf 721} (Oct., 2010) 939--955,
  [\href{http://xxx.lanl.gov/abs/1009.4191}{{\tt arXiv:1009.4191}}].

\bibitem{hahn10}
O.~{Hahn}, R.~{Teyssier}, and C.~M. {Carollo}, {\it {The large-scale
  orientations of disc galaxies}},  {\em \mnras} {\bf 405} (June, 2010)
  274--290, [\href{http://xxx.lanl.gov/abs/1002.1964}{{\tt arXiv:1002.1964}}].

\bibitem{bernstein02}
G.~M. {Bernstein} and M.~{Jarvis}, {\it {Shapes and Shears, Stars and Smears:
  Optimal Measurements for Weak Lensing}},  {\em \aj} {\bf 123} (Feb., 2002)
  583--618, [\href{http://xxx.lanl.gov/abs/astro-ph/0107431}{{\tt
  astro-ph/0107431}}].

\bibitem{smith2003}
R.~E. {Smith}, J.~A. {Peacock}, A.~{Jenkins}, S.~D.~M. {White}, C.~S. {Frenk},
  F.~R. {Pearce}, P.~A. {Thomas}, G.~{Efstathiou}, and H.~M.~P. {Couchman},
  {\it {Stable clustering, the halo model and non-linear cosmological power
  spectra}},  {\em \mnras} {\bf 341} (June, 2003) 1311--1332,
  [\href{http://xxx.lanl.gov/abs/astro-ph/0207664}{{\tt astro-ph/0207664}}].

\bibitem{pen00}
U.~{Pen}, J.~{Lee}, and U.~{Seljak}, {\it {Tentative Detection of Galaxy Spin
  Correlations in the Tully Catalog}},  {\em \apjl} {\bf 543} (Nov., 2000)
  L107--L110, [\href{http://xxx.lanl.gov/abs/astro-ph/0006118}{{\tt
  astro-ph/0006118}}].

\bibitem{mackey02}
J.~{Mackey}, M.~{White}, and M.~{Kamionkowski}, {\it {Theoretical estimates of
  intrinsic galaxy alignment}},  {\em \mnras} {\bf 332} (June, 2002) 788--798,
  [\href{http://xxx.lanl.gov/abs/astro-ph/0106364}{{\tt astro-ph/0106364}}].

\bibitem{mandelbaum10}
R.~{Mandelbaum} {\em et.~al.}, {\it {The WiggleZ Dark Energy Survey: direct
  constraints on blue galaxy intrinsic alignments at intermediate redshifts}},
  {\em \mnras} (Oct., 2010) 1486--+,
  [\href{http://xxx.lanl.gov/abs/0911.5347}{{\tt arXiv:0911.5347}}].

\bibitem{york00}
D.~G. {York} {\em et.~al.}, {\it {The Sloan Digital Sky Survey: Technical
  Summary}},  {\em \aj} {\bf 120} (Sept., 2000) 1579--1587,
  [\href{http://xxx.lanl.gov/abs/astro-ph/0006396}{{\tt astro-ph/0006396}}].

\bibitem{hirata04b}
C.~M. {Hirata} {\em et.~al.}, {\it {Galaxy-galaxy weak lensing in the Sloan
  Digital Sky Survey: intrinsic alignments and shear calibration errors}},
  {\em \mnras} {\bf 353} (Sept., 2004) 529--549,
  [\href{http://xxx.lanl.gov/abs/astro-ph/0403255}{{\tt astro-ph/0403255}}].

\bibitem{mandelbaum05}
R.~{Mandelbaum}, C.~M. {Hirata}, U.~{Seljak}, J.~{Guzik}, N.~{Padmanabhan},
  C.~{Blake}, M.~R. {Blanton}, R.~{Lupton}, and J.~{Brinkmann}, {\it
  {Systematic errors in weak lensing: application to SDSS galaxy-galaxy weak
  lensing}},  {\em \mnras} {\bf 361} (Aug., 2005) 1287--1322,
  [\href{http://xxx.lanl.gov/abs/astro-ph/0501201}{{\tt astro-ph/0501201}}].

\bibitem{blanton05}
M.~R. {Blanton} {\em et.~al.}, {\it {New York University Value-Added Galaxy
  Catalog: A Galaxy Catalog Based on New Public Surveys}},  {\em \aj} {\bf 129}
  (June, 2005) 2562--2578,
  [\href{http://xxx.lanl.gov/abs/astro-ph/0410166}{{\tt astro-ph/0410166}}].

\bibitem{kamionkowski98}
M.~{Kamionkowski}, A.~{Babul}, C.~M. {Cress}, and A.~{Refregier}, {\it {Theory
  and statistics of weak lensing from large-scale mass inhomogeneities}},  {\em
  \mnras} {\bf 301} (Dec., 1998) 1064--1072,
  [\href{http://xxx.lanl.gov/abs/astro-ph/9712030}{{\tt astro-ph/9712030}}].

\bibitem{crittenden02}
R.~G. {Crittenden}, P.~{Natarajan}, U.~{Pen}, and T.~{Theuns}, {\it
  {Discriminating Weak Lensing from Intrinsic Spin Correlations Using the
  Curl-Gradient Decomposition}},  {\em \apj} {\bf 568} (Mar., 2002) 20--27,
  [\href{http://xxx.lanl.gov/abs/astro-ph/0012336}{{\tt astro-ph/0012336}}].

\bibitem{zehavi10}
{The SDSS Collaboration}, I.~{Zehavi}, {\em et.~al.}, {\it {Galaxy Clustering
  in the Completed SDSS Redshift Survey: The Dependence on Color and
  Luminosity}},  {\em \rm ArXiv e-prints} (May, 2010)
  [\href{http://xxx.lanl.gov/abs/1005.2413}{{\tt arXiv:1005.2413}}].

\bibitem{macmillan06}
J.~D. {MacMillan}, L.~M. {Widrow}, and R.~N. {Henriksen}, {\it {On Universal
  Halos and the Radial Orbit Instability}},  {\em \apj} {\bf 653} (Dec., 2006)
  43--52, [\href{http://xxx.lanl.gov/abs/astro-ph/0604418}{{\tt
  astro-ph/0604418}}].

\bibitem{carpintero95}
D.~D. {Carpintero} and J.~C. {Muzzio}, {\it {Radial orbit instability in a
  Hubble-expanding universe}},  {\em \apj} {\bf 440} (Feb., 1995) 5--21.

\bibitem{bellovary08}
J.~M. {Bellovary}, J.~J. {Dalcanton}, A.~{Babul}, T.~R. {Quinn}, R.~W. {Maas},
  C.~G. {Austin}, L.~L.~R. {Williams}, and E.~I. {Barnes}, {\it {The Role of
  the Radial Orbit Instability in Dark Matter Halo Formation and Structure}},
  {\em \apj} {\bf 685} (Oct., 2008) 739--751,
  [\href{http://xxx.lanl.gov/abs/0806.3434}{{\tt arXiv:0806.3434}}].

\bibitem{kaiser87}
N.~{Kaiser}, {\it {Clustering in real space and in redshift space}},  {\em
  \mnras} {\bf 227} (July, 1987) 1--21.

\bibitem{tegmark06}
M.~{Tegmark} {\em et.~al.}, {\it {Cosmological constraints from the SDSS
  luminous red galaxies}},  {\em \prd} {\bf 74} (Dec., 2006) 123507,
  [\href{http://xxx.lanl.gov/abs/astro-ph/0608632}{{\tt astro-ph/0608632}}].

\bibitem{baldauf10}
T.~{Baldauf}, R.~E. {Smith}, U.~{Seljak}, and R.~{Mandelbaum}, {\it {Algorithm
  for the direct reconstruction of the dark matter correlation function from
  weak lensing and galaxy clustering}},  {\em \prd} {\bf 81} (Mar., 2010)
  063531, [\href{http://xxx.lanl.gov/abs/0911.4973}{{\tt arXiv:0911.4973}}].

\end{thebibliography}

\end{document}